\documentclass[aps,prb,superscriptaddress,two column,amsmath,amssymb]{revtex4-1}
\usepackage{amsmath}
\usepackage{graphicx}
\usepackage{color}
\usepackage{color}
\usepackage{graphicx}
\usepackage{natbib}
\usepackage{epsfig}
\usepackage{setspace}
\usepackage{amsmath,amssymb}
\usepackage{verbatim}
\usepackage{bibentry}
\usepackage{xcolor}
\usepackage{bm}
\usepackage{lipsum}
\usepackage{hyperref}

\begin{document}

	\title{Microwave-driven Floquet-Fano interference in a ring-chord quantum dot structure for enhanced spin-caloritronic performance}
	
\author{Parbati Senapati}
\affiliation{Theory Division, Saha Institute of Nuclear Physics, A CI of Homi Bhabha National Institute, Kolkata 700064, India}
\author{Kalpataru Pradhan}

\email{kalpataru.pradhan@saha.ac.in}

\affiliation{Theory Division, Saha Institute of Nuclear Physics, A CI of Homi Bhabha National Institute, Kolkata 700064, India}

\begin{abstract}

We investigate photon-assisted thermoelectric transport in four--quantum-dot nanostructures
featuring ring and ring--chord geometries coupled to ferromagnetic leads. Focusing on the interplay
between microwave-induced Floquet sidebands and geometry-driven Fano interference, we employ
the nonequilibrium Green's function formalism combined with Floquet theory and a self-consistent
Hartree treatment of electron-electron interactions within the linear response regime. The inclusion
of a longitudinal interdot chord bridging the lead-coupled dots introduces a discrete interference
pathway that competes with the continuum of ring-mediated states, giving rise to pronounced Fano
resonances. Microwave irradiation dynamically reshapes these resonances through photon
absorption and emission processes, enabling tunable control of electrical conductance,
thermopower, and electronic thermal conductance. Quantitatively, at an intermediate temperature of $T=0.3\Gamma_0$ (where $\Gamma_0$ denotes the dot-lead coupling strength), the microwave-driven ring-chord geometry exhibits an exceptional thermoelectric figure of merit of ZT $\approx$ 12 and an optimal efficiency--power trade-off, reaching nearly $62\%$ of the Carnot efficiency with an output power of $6.24~\mathrm{fW}$. Crucially, the combination of spin-polarized
injection from the leads and Zeeman splitting within the dots induces a robust spin-dependence
within this Floquet--Fano interference. This cooperative interplay results in an enhanced spin
Seebeck response and a maximum spin thermoelectric figure of merit of nearly $Z_sT \approx 18$. Our findings establish microwave-driven engineering of Fano interference as an effective strategy for modulating spin-caloritronic behavior in multi-quantum-dot devices.
\end{abstract}
\maketitle
\section{Introduction}

Thermoelectric transport in nanoscale quantum systems has remained a focal point of research due to its fundamental importance and its potential applications in energy harvesting, solid-state cooling, and spin-based information processing
\cite{minnich2009bulk,galperin2007molecular,zhao2023quantum,dubi2011colloquium,wierzbicki2010electric,trocha2012large}.
In low-dimensional nanostructures, thermoelectric response is governed not only by material properties but also by quantum confinement, strong electron--electron interactions, and phase-coherent transport\cite{senapati2025thermoelectric,behera2023quantum,zheng2012thermoelectric,swirkowicz2009thermoelectric,scheibner2005thermopower,liu2010enhancement}.
These features allow thermoelectric behavior to be engineered at the level of individual quantum states, thereby offering possibilities that are inaccessible in bulk systems.

A particularly powerful route to controlling thermoelectric transport at the nanoscale is provided by external time-periodic driving.
When quantum conductors are subjected to microwave or ac electric fields, they are driven into nonequilibrium steady states in which electrons can exchange energy quanta with the driving field
\cite{dubi2009thermospin,swirkowicz2009thermoelectric,liu2010enhancement}.
In this regime, transport is naturally described within the Floquet formalism, where electronic states become dressed by photons, effectively reconfiguring the energy spectrum into a ladder of sidebands separated by integer multiples of the driving frequency
\cite{crepieux2011enhanced,chi2012microwave,bagheri2013thermoelectric,chi2012microwave}.
These photon-assisted processes fundamentally reshape the transmission function, which governs electrical conductance, thermopower, and thermal conductance.

Quantum dots (QDs) provide a versatile platform for exploring photon-assisted thermoelectric effects.
Their discrete energy spectra, strong Coulomb interaction, and highly tunable tunnel couplings allow for systematic control of transport pathways under microwave irradiation.
Within the Floquet picture, photon-assisted tunneling enables electrons to absorb or emit energy quanta while traversing the device, opening additional transport channels and producing characteristic sideband structures in the transmission spectrum
\cite{mosallanejad2024floquet,zhang2025enhanced,abdullah2019photon}.
Importantly, microwave fields act as nonthermal control parameters, avoiding the need to modify static gate voltages or reservoir temperatures to achieve dynamic tuning of thermoelectric properties
\cite{bajpai2020robustness,platero2004photon,de2024semi}.

When a quantum dot is coupled to ferromagnetic leads, photon-assisted thermoelectric transport becomes intrinsically spin dependent due to the spin-polarized reservoirs. The exchange-split density of states induces asymmetric coupling for spin-up and spin-down channels, which is directly inherited by the Floquet sidebands under microwave driving
\cite{mosallanejad2024floquet}.
This asymmetry stimulate the emergence of spin-dependent transmission, spin thermopower, finite spin Seebeck effects, and enhanced spin thermoelectric figures of merit
\cite{trocha2025spin,wrzesniewski2025effects}.
As a result, microwave-driven quantum-dot systems provide a natural setting for investigating nonequilibrium spin caloritronics within a fully quantum-coherent framework. Liu et al. \cite{liu2012proposal} demonstrated that a single-level quantum dot driven by a time-dependent magnetic field can act as a source of spin current, with the thermopower exhibiting a pronounced enhancement in the transient regime. Furthermore, recent reports \cite{tagani2014time,crepieux2011enhanced} indicate a rapid rise in thermopower under a step-like gate voltage modulation. In a related study, Chi and Dubi investigated the thermoelectric response of a quantum dot coupled to metallic leads under microwave irradiation,  finding that the application of microwave fields substantially enrich the resonant landscape and enhance the overall thermoelectric figure of merit \cite{chi2012microwave}.

The interplay between Floquet physics and quantum interference becomes particularly rich in systems composed of multiple coupled quantum dots.
In such systems, photon-assisted sidebands hybridize with molecular-like states arising from interdot tunneling, giving rise to highly structured energy-dependent transmission spectra
\cite{giesen2025tunneling,tsuji2023floquet}.
Moving beyond from linear dot chains to closed geometries significantly enlarges the landscape of coherent transport pathways and interference phenomena.
In particular, four--quantum-dot configurations support ring geometries in which electrons can circulate along multiple phase-coherent paths, as well as additional chord-like tunneling links that directly connect the dots coupled to the leads (see Fig. \ref{schematic}).

The presence of a chord in a four--quantum-dot ring introduces a discrete interference channel that competes with the quasi-continuum of ring-mediated pathways.
This competition naturally gives rise to Fano-type interference, characterized by asymmetric line shapes and antiresonances in the transmission spectrum.
Under microwave irradiation, these Fano resonances acquire a Floquet character, as photon-assisted sidebands of both the discrete and continuum states participate in transport\cite{wang2016resonant,zhao2015fano}.
The resulting photon-assisted Fano interference provides a highly sensitive mechanism for controlling thermoelectric response, facilitating sharp modulation of conductance, thermopower, and thermal conductance through small changes in driving parameters.

To capture these effects quantitatively, a theoretical framework capable of treating time-periodic driving, quantum coherence, and interactions on equal footing is required.
The nonequilibrium Green's function (NEGF) formalism, combined with Floquet theory, provides such a framework and stand as the standard tool for analyzing photon-assisted transport in driven nanostructures.
Within this approach, the full energy-dependent transmission function--including Floquet sidebands, interference effects and interaction induced level shifts--can be computed self-consistently. This enables the rigorous/direct evaluation of thermoelectric transport coefficients in the linear-response regime.

Motivated by these considerations, we investigate photon-assisted thermoelectric transport in four--quantum-dot nanostructures with ring and ring--chord geometries coupled to ferromagnetic leads.
The organization of the paper is as follows. In Sec.~\ref{sec:model}, we introduce the theoretical model and formalism, providing a unified framework for examining the combined effects of microwave driving, strong Coulomb interaction, spin polarization, and geometry-induced Fano interference on thermoelectric properties.
Employing a Floquet-based nonequilibrium Green's function approach with a self-consistent Hartree treatment of electron--electron interactions, we calculate spin-resolved charge and spin thermoelectric transport coefficients. We further analyze their sensitivity to microwave amplitude and frequency, temperature, and geometric configuration.
The results of our numerical analysis and the correcsponding discussion are presented in Sec.~\ref{sec:results}, followed by our conclusions in Sec.~\ref{sec:conclusions}.
\section{Model and theoretical formalism}
\label{sec:model}

\begin{figure}[h!]
	\centering
	\includegraphics[width=1.0\columnwidth]{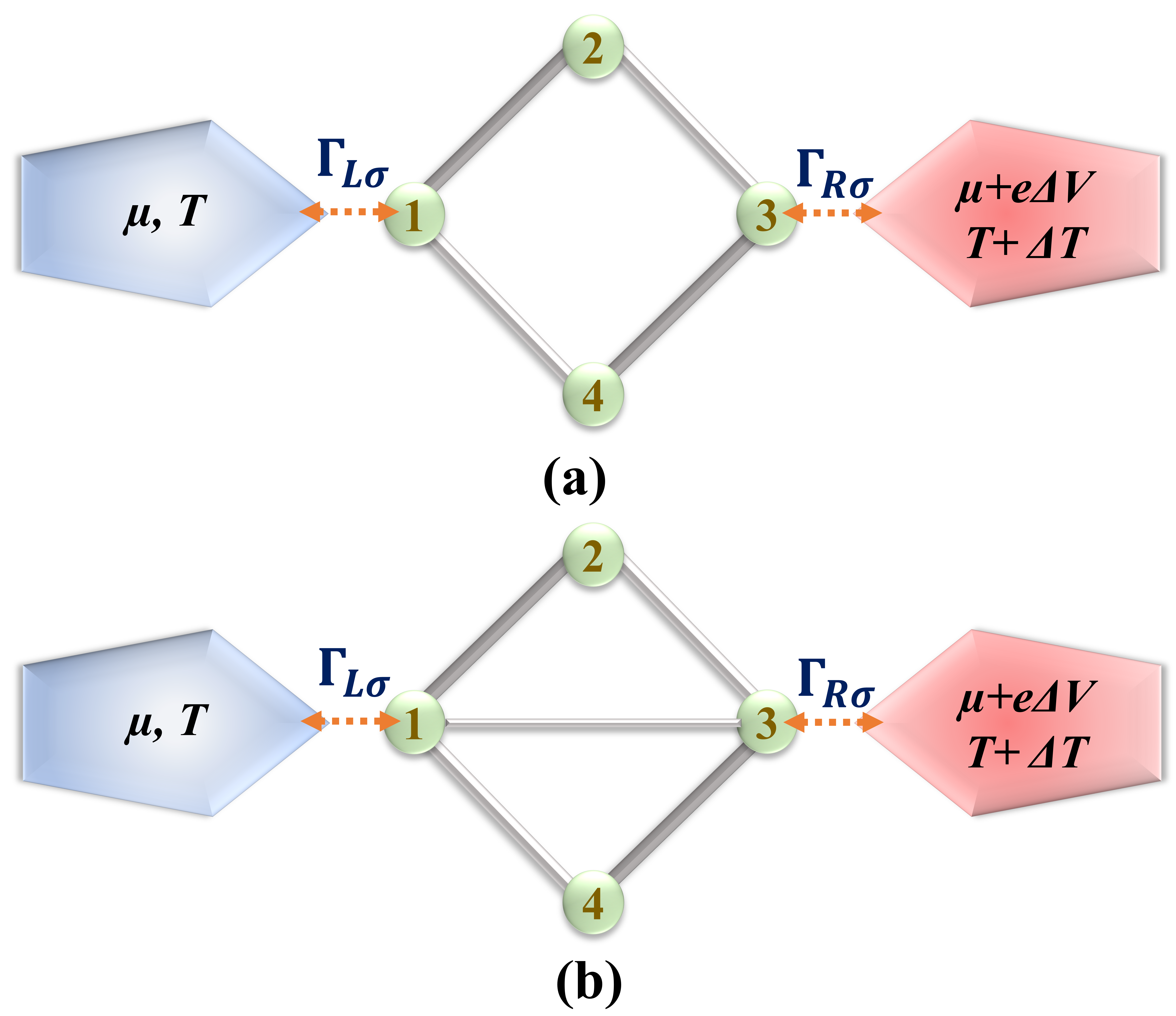}
	\caption{\label{schematic}Schematic illustration of the four--quantum-dot nanostructure coupled to ferromagnetic electrodes: (a) Pure ring geometry, where four quantum dots form a closed loop and are connected to the left and right leads through dots 1 and 3 with spin-dependent tunneling rates $\Gamma_{L\sigma}$ and $\Gamma_{R\sigma}$, respectively. (b) Ring--chord geometry, where an additional direct coupling (chord) between dots 1 and 3 introduces an alternative transport pathway.}
	\end{figure}

We consider a system consisting of four single-level quantum dots arranged in ring and ring–chord geometries, coupled to two ferromagnetic leads and driven by a time-periodic microwave field, as schematically illustrated in Fig.~\ref{schematic}. The system is treated within the nonequilibrium Green's function (NEGF) formalism combined with Floquet theory in order to account for photon-assisted transport \cite{wingreen1993time,sun1997time,meir1992landauer,liu2010enhancement,haug2008quantum}. Electron--electron interaction on each quantum dot is treated self-consistently at the Hartree level. Throughout this work we use units such that $\hbar = k_B = e = 1$. 

The total Hamiltonian is written as
\begin{equation}
H(t) = H_{\mathrm{leads}} + H_{\mathrm{dots}}(t) + H_{\mathrm{T}},
\label{eq:Htot}
\end{equation}
where $H_{\mathrm{leads}}$ describes the ferromagnetic reservoirs, $H_{\mathrm{dots}}(t)$ represents the interacting quantum dots under microwave driving, and $H_{\mathrm{T}}$ is the tunnel coupling between the dots and the leads.

The ferromagnetic leads are modeled as noninteracting electron reservoirs,
\begin{equation}
H_{\mathrm{leads}} =
\sum_{\alpha=L,R}
\sum_{k,\sigma}
\varepsilon_{\alpha k\sigma}
c^{\dagger}_{\alpha k\sigma}
c_{\alpha k\sigma},
\label{eq:Hleads}
\end{equation}
where $c^{\dagger}_{\alpha k\sigma}$ ($c_{\alpha k\sigma}$) creates (annihilates) an electron with momentum $k$ and spin $\sigma=\uparrow,\downarrow$ in lead $\alpha$. Spin polarization of the leads is incorporated through spin-dependent tunnel couplings,
\begin{equation}
\Gamma_{\alpha\sigma} = \Gamma_0 (1 + \sigma q),
\label{eq:Gamma}
\end{equation}
where $\Gamma_0$ denotes the average coupling strength and $q$ is the polarization parameter.

The Hamiltonian of the four coupled quantum dots is
\begin{align}
H_{\mathrm{dots}}(t) =
&\sum_{i=1}^{4}
\sum_{\sigma}
\varepsilon_{i\sigma}(t)\,
d^{\dagger}_{i\sigma} d_{i\sigma}
+
U \sum_{i=1}^{4}
n_{i\uparrow} n_{i\downarrow}
\nonumber\\
&-
\sum_{\langle i,j\rangle}
\sum_{\sigma}
\left(
t_{ij}\,
d^{\dagger}_{i\sigma} d_{j\sigma}
+
\mathrm{H.c.}
\right),
\label{eq:Hdots}
\end{align}
where $d^{\dagger}_{i\sigma}$ creates an electron with spin $\sigma$ on dot $i$,
$n_{i\sigma}=d^{\dagger}_{i\sigma}d_{i\sigma}$ is the number operator, and $U$ is the on-site Coulomb interaction.
Nearest-neighbor hoppings $t_{12}$, $t_{23}$, $t_{34}$, and $t_{41}$ describe electron tunneling between quantum dots $(1 \leftrightarrow 2)$, $(2 \leftrightarrow 3)$, $(3 \leftrightarrow 4)$, and $(4 \leftrightarrow 1)$, respectively, thereby forming a closed ring geometry. In addition, the optional chord hopping $t_{13}$ represents a direct tunneling path between dots $(1 \leftrightarrow 3)$.

The microwave field is introduced as a harmonic modulation of the dot energy levels:
\begin{equation}
\varepsilon_{i\sigma}(t)
=
\varepsilon_i + \sigma E_Z + \Delta_d \cos(\omega t),
\label{eq:microwave}
\end{equation}
where $\varepsilon_i$ are the static dot energies, $E_Z$ is the Zeeman splitting, and $\Delta_d$ and $\omega$ are the amplitude and angular frequency of the microwave field, respectively.

The tunnel coupling between the quantum dots and the leads is described by
\begin{equation}
H_{\mathrm{T}} =
\sum_{k,\sigma}
\left(
V_{Lk\sigma}\,
c^{\dagger}_{Lk\sigma} d_{1\sigma}
+
V_{Rk\sigma}\,
c^{\dagger}_{Rk\sigma} d_{3\sigma}
+
\mathrm{H.c.}
\right),
\label{eq:HT}
\end{equation}
such that the left (right) lead couples exclusively to dot~1 (dot~3).
Within the wide-band approximation, the retarded self-energies due to the leads are purely imaginary and take the form
\begin{equation}
\boldsymbol{\Sigma}^{r}_{L\sigma}
=
\begin{pmatrix}
-\frac{i}{2}\Gamma_{L\sigma} & 0 & 0 & 0 \\
0 & 0 & 0 & 0 \\
0 & 0 & 0 & 0 \\
0 & 0 & 0 & 0
\end{pmatrix},
\qquad
\boldsymbol{\Sigma}^{r}_{R\sigma}
=
\begin{pmatrix}
0 & 0 & 0 & 0 \\
0 & 0 & 0 & 0 \\
0 & 0 & -\frac{i}{2}\Gamma_{R\sigma} & 0 \\
0 & 0 & 0 & 0
\end{pmatrix}.
\label{eq:selfenergy}
\end{equation}

Electron--electron interaction is treated at the Hartree level, leading to effective spin-dependent dot energies
\begin{equation}
\varepsilon^{\mathrm{eff}}_{i\sigma}
=
\varepsilon_i + \sigma E_Z + U \langle n_{i\bar{\sigma}} \rangle,
\label{eq:Hartree}
\end{equation}
where the occupations $\langle n_{i\sigma} \rangle$ are determined self-consistently from the local spectral functions.
The corresponding effective single-particle Hamiltonian matrix reads
\begin{equation}
\mathbf{H}^{\mathrm{eff}}_{\sigma}
=
\begin{pmatrix}
\varepsilon^{\mathrm{eff}}_{1\sigma} & -t_{12} & -t_{13} & -t_{41} \\
-t_{12} & \varepsilon^{\mathrm{eff}}_{2\sigma} & -t_{23} & 0 \\
-t_{13} & -t_{23} & \varepsilon^{\mathrm{eff}}_{3\sigma} & -t_{34} \\
-t_{41} & 0 & -t_{34} & \varepsilon^{\mathrm{eff}}_{4\sigma}
\end{pmatrix}.
\label{eq:Heff_matrix}
\end{equation}

Because the Hamiltonian is periodic in time, $H(t+T)=H(t)$ with $T=2\pi/\omega$, the system can be treated using Floquet theory.
The retarded Green's function for spin $\sigma$ and photon sideband index $k$ is given by
\begin{equation}
\mathbf{G}^{r}_{\sigma}(E,k)
=
\left[
(E + k\omega)\mathbf{I}
-
\mathbf{H}^{\mathrm{eff}}_{\sigma}
-
\boldsymbol{\Sigma}^{r}_{L\sigma}
-
\boldsymbol{\Sigma}^{r}_{R\sigma}
\right]^{-1},
\label{eq:Gr}
\end{equation}
where the index $k$ labels photon sidebands associated with the absorption ($k>0$) or emission ($k<0$) of $|k|$ photons from the microwave field.

The spin-resolved transmission function is obtained by summing over all photon sidebands,
\begin{equation}
T_{\sigma}(E)
=
\sum_{k=-N_{\mathrm{ph}}}^{N_{\mathrm{ph}}}
J_k^2\!\left(\frac{\Delta_d}{\omega}\right)
\Gamma_{L\sigma}\Gamma_{R\sigma}
\left|
G^{r}_{13,\sigma}(E,k)
\right|^2,
\label{eq:Transmission}
\end{equation}
where $J_k$ is the Bessel function of the first kind and $N_{\mathrm{ph}}$ denotes the Floquet truncation order.

Within linear-response theory, thermoelectric transport coefficients are expressed in terms of the moments
\begin{equation}
L_{n\sigma}
=
\frac{1}{\pi}
\int dE\,
(E-\mu)^n
\left(-\frac{\partial f}{\partial E}\right)
\mathcal{T}_{\sigma}(E),
\label{eq:Ln}
\end{equation}
with $f(E)$ the Fermi--Dirac distribution and $\mu$ the chemical potential.

The electrical conductance is given by
\begin{equation}
G_{\sigma} = L_{0\sigma}, \qquad
G = G_{\uparrow} + G_{\downarrow},
\label{eq:G}
\end{equation}
while the Seebeck coefficient (thermopower) for each spin channel reads
\begin{equation}
S_{\sigma} = -\frac{L_{1\sigma}}{T L_{0\sigma}}.
\label{eq:Ssigma}
\end{equation}
The total charge and spin thermopowers are defined as
\begin{equation}
S = S_{\uparrow} + S_{\downarrow}, \qquad
S_s = S_{\uparrow} - S_{\downarrow}.
\label{eq:SchargeSpin}
\end{equation}

To characterize the heat transport, the electronic thermal conductance is
\begin{equation}
\kappa_{\sigma}
=
\frac{1}{T}
\left(
L_{2\sigma}
-
\frac{L_{1\sigma}^2}{L_{0\sigma}}
\right),
\qquad
\kappa = \kappa_{\uparrow} + \kappa_{\downarrow}.
\label{eq:kappa}
\end{equation}

The dimensionless charge and spin thermoelectric figures of merit are defined as:
\begin{equation}
ZT
=
\frac{S^2GT}{\kappa} 
=\frac{
	\left(L_{1\sigma}\right)^2
}
{
	\left(L_{0\sigma}\right)
	\left(L_{2\sigma}\right)
	-
	\left(L_{1\sigma}\right)^2
},
\label{eq:ZT}
\end{equation}

\begin{equation}
Z_sT
= \frac{|G_s|S_s^2T}{\kappa}
=\frac{
	\left(L_{1\uparrow}-L_{1\downarrow}\right)^2
}
{
	\left|L_{0\uparrow}-L_{0\downarrow}\right|
	\left(L_{2\uparrow}+L_{2\downarrow}\right)
	-
	\left(L_{1\uparrow}-L_{1\downarrow}\right)^2
},
\label{eq:ZsT}
\end{equation}

with $G_s = G_{\uparrow} - G_{\downarrow}$. In the above expressions, $\kappa$ denotes the electronic contribution to the thermal conductance, while the phonon contribution is neglected.
For quantum dot nanostructures such as the four--quantum-dot ring--chord system considered here, phonon heat transport is typically limited by weak mechanical coupling and strong vibrational mismatch at the dot--lead tunneling interfaces.
Consequently, the thermal conductance is dominated by its electronic component, and the charge and spin thermoelectric figures of merit are evaluated using the electronic thermal conductance alone. Therefore, when comparing with experimental results, the qualitative trends obtained here are expected to be relevant, whereas the precise numerical values of the dimensionless figure of merit may differ from those measured in realistic devices \cite{liu2009thermoelectric,bai2018influence,khedri2017influence}.

The efficiency at maximum power relative to the Carnot efficiency is evaluated as:\cite{whitney2014most,whitney2015finding,nakpathomkun2010thermoelectric}
\begin{equation}
\frac{\eta}{\eta_C}
=
\frac{\sqrt{1+ZT}-1}{\sqrt{1+ZT}+\frac{T_c}{T_h}},
\label{eq:efficiency}
\end{equation}
where $T_c$ and $T_h$ are the temperatures of the cold and hot leads, respectively.

The maximum output power generated by the thermoelectric device is expressed as
\begin{equation}
P_{\mathrm{max}}
=
\frac{1}{4} \, G S^2 (\Delta T)^2,
\label{eq:maxpower}
\end{equation}
where $\Delta T = T_h-T_c$ is the applied temperature difference between the two leads. Since the quantity $GS^2$ corresponds to the power factor, the maximum power output is directly governed by the combined electrical and thermoelectric transport properties of the system.

This formalism enables a direct and systematic comparison of microwave-driven thermoelectric transport in four--quantum-dot systems with different geometrical configurations, including closed ring structures and ring geometries with an additional interdot chord, under identical interaction, driving, and lead-coupling conditions.

We set the parameter $\Gamma_0=1.0$, which represents the lead--dot coupling strength and use it as the reference energy scale throughout the calculations. Accordingly, all energy-related parameters, including the hopping amplitudes $(t_{ij})$, temperature $(T)$, quantum dot energy levels $(\varepsilon_i)$, onsite Coulomb interaction $(U)$, Zeeman splitting $(E_z)$, and microwave field strength $(\Delta_d)$, are expressed in units of $\Gamma_0$. Unless otherwise stated, all calculations are performed at temperature $T = 0.3$.

\section{Results and discussion}
\label{sec:results}
\subsection{Zero-microwave transmission and Fano interference}

\begin{figure}[h!]
	\centering
	\includegraphics[width=1.0\columnwidth]{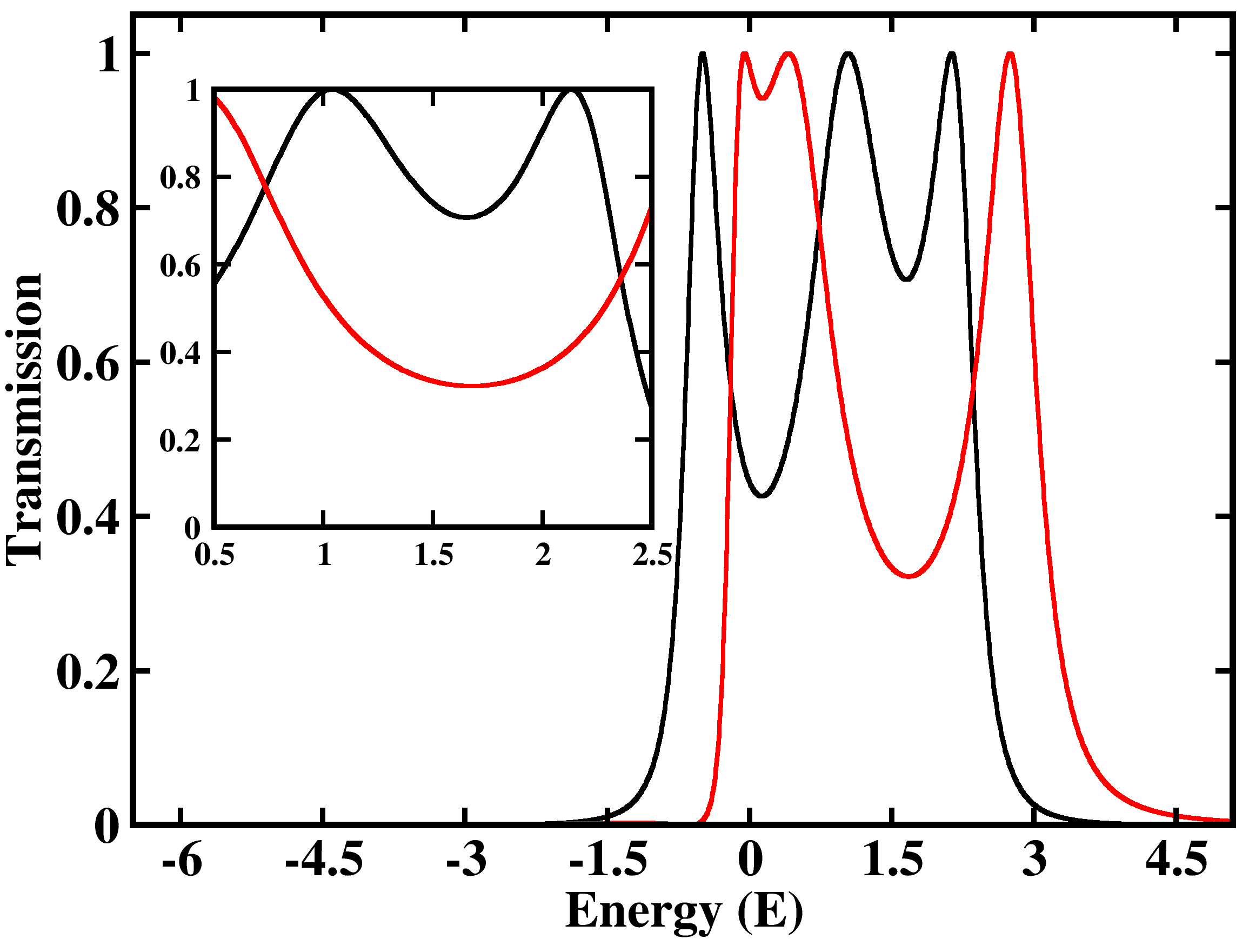}
	\caption{\label{Transmission}Zero-microwave transmission $\mathcal{T}(E)$ as a function of energy for the four--quantum-dot system, where the black curve corresponds to the closed ring geometry ($t_{13}=0$) and the red curve to the ring--chord configuration ($t_{13}\neq0$), showing geometry-induced Fano interference. Inset: Magnified view of the central energy window $E\in[0.5,\,2.5]$ highlighting the asymmetric line shape and antiresonance-like suppression of transmission in the ring--chord case.}
	
\end{figure}

We begin by examining the energy-dependent transmission in the absence of microwave driving ($\Delta_d=0$), which provides direct insight into geometry-induced quantum interference in the four--quantum-dot system.  The transmission function $\mathcal{T}(E)$ is shown in Fig.~\ref{Transmission}. The black curve corresponds to the closed ring configuration with $t_{13}=0$, whereas the red curve represents the ring--chord configuration with $t_{13}\neq0$. The nearest-neighbor hopping parameters are chosen as $t_{12}=t_{23}=t_{34}=t_{41}=0.7$, while the on-site Coulomb interaction and temperature are taken as $U=2$ and $T=0.3$, respectively. 

For the pure ring geometry, the transmission exhibits quasi-symmetric resonant features about the band center ($E \approx 0$). This behavior originates from the approximate particle--hole (chiral) symmetry of the system: for a symmetric four-site ring with uniform on-site energies, the Hamiltonian (after shifting by the on-site energy) satisfies $\mathcal{C} H \mathcal{C}^{-1} = -H$, with $\mathcal{C}=\mathrm{diag}(1,-1,1,-1)$ denoting the chiral operator, leading to eigenvalue pairs $\pm E$ and consequently $\mathcal{T}(E) \approx -\mathcal{T}(E)$. The quasi-symmetry is further reflected in corresponding transmission peaks at energies $\pm E$ (e.g., near $\pm 1.4$), which exhibit comparable heights and widths, with small deviations arising from the energy-dependent lead self-energies that break exact particle--hole symmetry. These features indicate transport through extended eigenstates of the ring, formed by coherent superpositions of the four quantum-dot levels coupled via uniform interdot hopping. In the absence of the chord, transport proceeds exclusively through these ring-mediated pathways and is therefore governed by conventional resonant tunneling.

Introducing the interdot chord between dots 1 and 3 opens an additional direct propagation channel that competes coherently with the ring-mediated transport.
Under the present symmetric coupling conditions, the chord effectively acts as a discrete shortcut connecting the lead-coupled dots, while the ring continues to provide a continuum-like background of extended states.
As a result, the total transmission amplitude becomes a coherent superposition of chord-assisted and ring-assisted contributions, giving rise to Fano-type quantum interference.

This interference manifests itself in Fig.~\ref{Transmission} through two characteristic features.
First, the transmission in the ring--chord geometry (red curve) displays strongly asymmetric line shapes that deviate markedly from the quasi-symmetric resonances observed for the pure ring. Most notably, the introduction of the chord significantly alters the interference landscape: the prominent transmission minimum observed for the pure ring at E=0 almost vanishes, whereas a much deeper and more pronounced supression develpos in the region around E$\approx$1.5.
Second, and more importantly, the chord induces a pronounced suppression of transmission in an energy range where the ring-only transmission remains large.
In particular, within the central main transport window in the pure ring case (approximately $E\in[0.5,\,2.5]$ in the absolute energy units of the figure), the red curve develops a significant dip relative to the black curve.
This antiresonance-like suppression originates from destructive interference between the chord-assisted shortcut and the ring-mediated propagation paths, whose relative phase leads to partial cancellation of the transmission amplitude.

We emphasize that this suppression occurs without invoking any magnetic flux and for moderate temperature $T=0.3$, demonstrating that the observed Fano interference is a purely geometry-driven effect and remains robust against thermal broadening.
Accordingly, Fig.~\ref{Transmission} establishes that the addition of the chord qualitatively transforms the four--quantum-dot ring into a Fano interferometer. This modification suppresses transmission within the central transport window via destructive interference, thereby providing the foundation for the photon-assisted and thermoelectric interference effects discussed in the following sections.

\subsection{Electrical and thermal conductance under microwave driving}

\begin{figure}[h!]
	\centering
	\includegraphics[width=1.0\columnwidth]{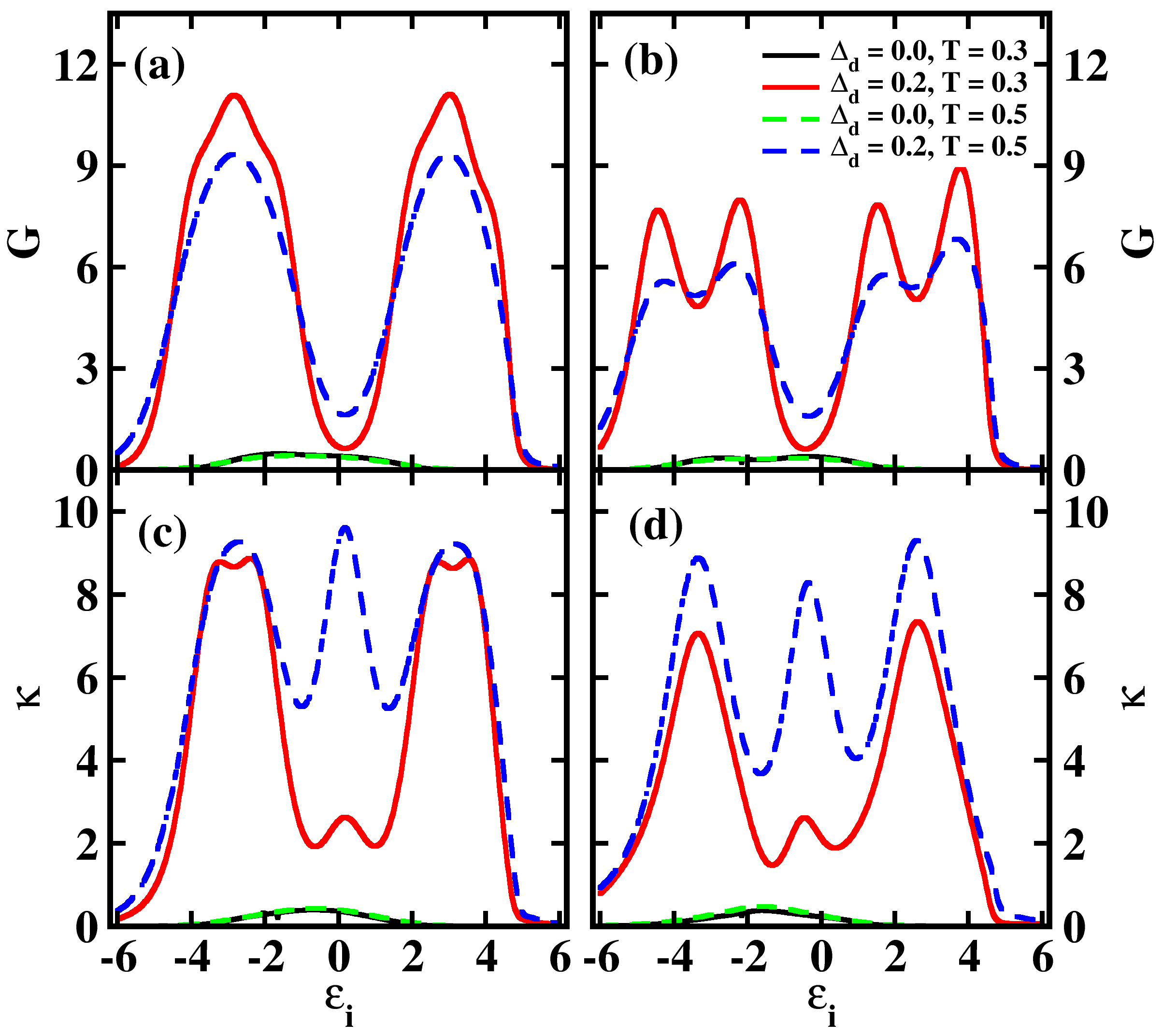}
	\caption{\label{gv_ke}{Electrical conductance $G$ in panels (a) and (b), and thermal conductance $\kappa$ in panels (c) and (d), plotted as functions of the normalized on-site energy $\varepsilon_i/\Gamma_0$ for both undriven and driven cases at two different temperatures. Panels (a) and (c) correspond to the pure ring configuration, whereas panels (b) and (d) represent the ring--chord geometry.}}
\end{figure}

Having established the presence of Fano interference at the level of energy-resolved transmission, we now investigate how this interference is reflected in relevant transport coefficients in the presence of microwave driving. We present the electrical conductance $G$ [panels (a) and (b)] and the electronic thermal conductance $\kappa$ [panels (c) and (d)] as functions of the normalized on-site energy $\varepsilon_i/\Gamma_0$ for the pure ring geometry (left column) and the ring--chord geometry (right column), as shown in Fig.~\ref{gv_ke}.
The black (red) solid curves correspond to the undriven ($\Delta_d = 0$) [driven ($\Delta_d = 0.2$)] case at $T = 0.3$, while the green (blue) dashed curves denote the corresponding cases at $T = 0.5$.
All other parameters are identical to those used in the transmission analysis, and all quantities are expressed in the same energy units.

For the pure ring configuration, both $G$ and $\kappa$ exhibit relatively smooth and broad variations when the microwave field is absent.
This behavior is characteristic of transport dominated by extended ring eigenstates, where resonant tunneling leads to moderate conductance peaks and only weak suppression of heat flow.
When the microwave field is applied, photon-assisted tunneling activates additional Floquet sidebands, which substantially enhance the electrical conductance over a wide range of $\varepsilon_i$.
At the same time, the thermal conductance is also increased, reflecting the participation of additional energy-carrying channels.
In this geometry, microwave driving primarily redistributes spectral weight without introducing strong interference-based filtering between charge and heat transport.

In contrast, the ring--chord geometry exhibits a qualitatively different response.
Even in the absence of microwave driving, both $G$ and $\kappa$ display stronger modulation compared with the pure ring, reflecting geometry-induced Fano interference between the chord-assisted shortcut and the ring-mediated propagation paths.
This interference selectively suppresses transport in specific energy regions. The effect is particularly pronounced in the thermal conductance, indicating that heat carrying electrons are more efficiently filtered than charge carrying one.

The impact of microwave driving is further amplified in the ring--chord system.
When $\Delta_d$ is finite, the combined action of Floquet sidebands and Fano interference produces a highly structured response. In the ring-chord geometry, while the absolute magnitude of $G$ is lower than in the pure ring, it maintains a robust multi-peak profile. In contrast, the thermal conductance $\kappa$ exhibits a more constricted structure. While the electrical conductance $G$ maintains a broad profile with significant spectral weight between the sidebands, the peaks in $\kappa$ are noticeably narrower. This indicates that the Fano interference more aggressively truncates the energy channels responsible for heat transport, effectively thinning the thermal response while leaving electrical response more robust. 
This imbalance between charge and heat transport is considerably more significant than in the pure ring geometry, demonstrating that microwave driving acts most effectively when applied to an interference-optimized structure.

An increase in temperature leads to a noticeable suppression and broadening of the electrical conductance $G$ in both geometries for $\Delta_d=0.2$, with the effect being more pronounced in the ring--chord configuration. In the pure ring geometry, the peaks in $G$ decrease moderately in height and become smoother while largely preserving their symmetric structure about $\varepsilon_i = 0$. In contrast, the ring--chord geometry exhibits stronger thermal smearing, where multiple resonant peaks merge and lose sharpness, reflecting enhanced sensitivity of quantum interference to temperature. For $\Delta_d=0$, however, the electrical conductance remains nearly unaffected, with only very small changes observed upon increasing temperature. In the case of thermal conductance $\kappa$, the behavior is qualitatively different: $\kappa$ increases with temperature in both geometries due to the broader energy window of thermally activated carriers contributing to heat transport. For the pure ring, this increase occurs gradually while maintaining an overall symmetric profile, whereas in the ring--chord geometry, the enhancement in $\kappa$ is more pronounced and accompanied by significant peak broadening and partial filling of minima. In contrast, for $\Delta_d=0$, the thermal conductance also shows only weak variation with temperature. The modification in $\kappa$ directly influences the thermoelectric figure of merit $ZT$, which will be discussed in detail later. Consequently, while increasing temperature weakens the coherent resonant features in $G$, it simultaneously enhances $\kappa$, with the ring--chord system exhibiting a stronger temperature dependence in both quantities due to its additional interference pathways.

The connection between Fig.~\ref{Transmission} and Fig.~\ref{gv_ke} becomes clear when recalling that the transport coefficients are energy-weighted integrals of the transmission, with $G\propto L_0$ and $\kappa \propto L_2 - L_1^2/L_0$ [Eqs.~(\ref{eq:Ln})--(\ref{eq:kappa})], where the weighting function $-\partial f/\partial E$ restricts the dominant contributions to an energy window of width $\sim T$ (with $k_B = 1$) around the chemical potential.
When the gate-controlled level position $\varepsilon_i$ is tuned such that this Fermi window overlaps the Fano-suppressed region of $\mathcal{T}(E)$ identified in Fig.~\ref{Transmission}, the ring--chord device exhibits a pronounced reduction and reshaping of the transport integrals.
Because $\kappa$ involves higher energy moments of the transmission, it is more sensitive to the sharp peak--dip structures associated with Fano interference, leading to a stronger suppression of thermal conductance than electrical conductance.
The more effective suppression of $\kappa$ relative to $G$ in the ring-chord geometry fulfills a primary goal of thermoelectrics; it suggests that, this geometry can maintain electrical flow while blocking heat, which is the key to achieving a high thermoelectric figure of merit (ZT).

Finally, the disintegrated multi-peak structure in the conductance can be understood as a consequence of interference-induced spectral filtering combined with thermal broadening.
In the ring--chord geometry, destructive interference suppresses transmission over extended energy intervals, fragmenting the transport into series of narrow resonances or distinct photon-assisted sidebands when microwave driving is present.
When this surviving resonance is aligned with the chemical potential by tuning $\varepsilon_i$, a single pronounced maximum emerges in $G(\varepsilon_i)$, while neighboring resonances are either quenched by Fano antiresonances or shifted outside the Fermi window.
This fragmented peak behavior thus provides a clear signature of the chord acting as an interference-based energy filter, simultaneously sharpening the electrical response and suppressing heat transport.

\subsection{Seebeck coefficient and figure of merit under microwave driving}

\begin{figure}[h!]
	\centering
	\includegraphics[width=1.0\columnwidth]{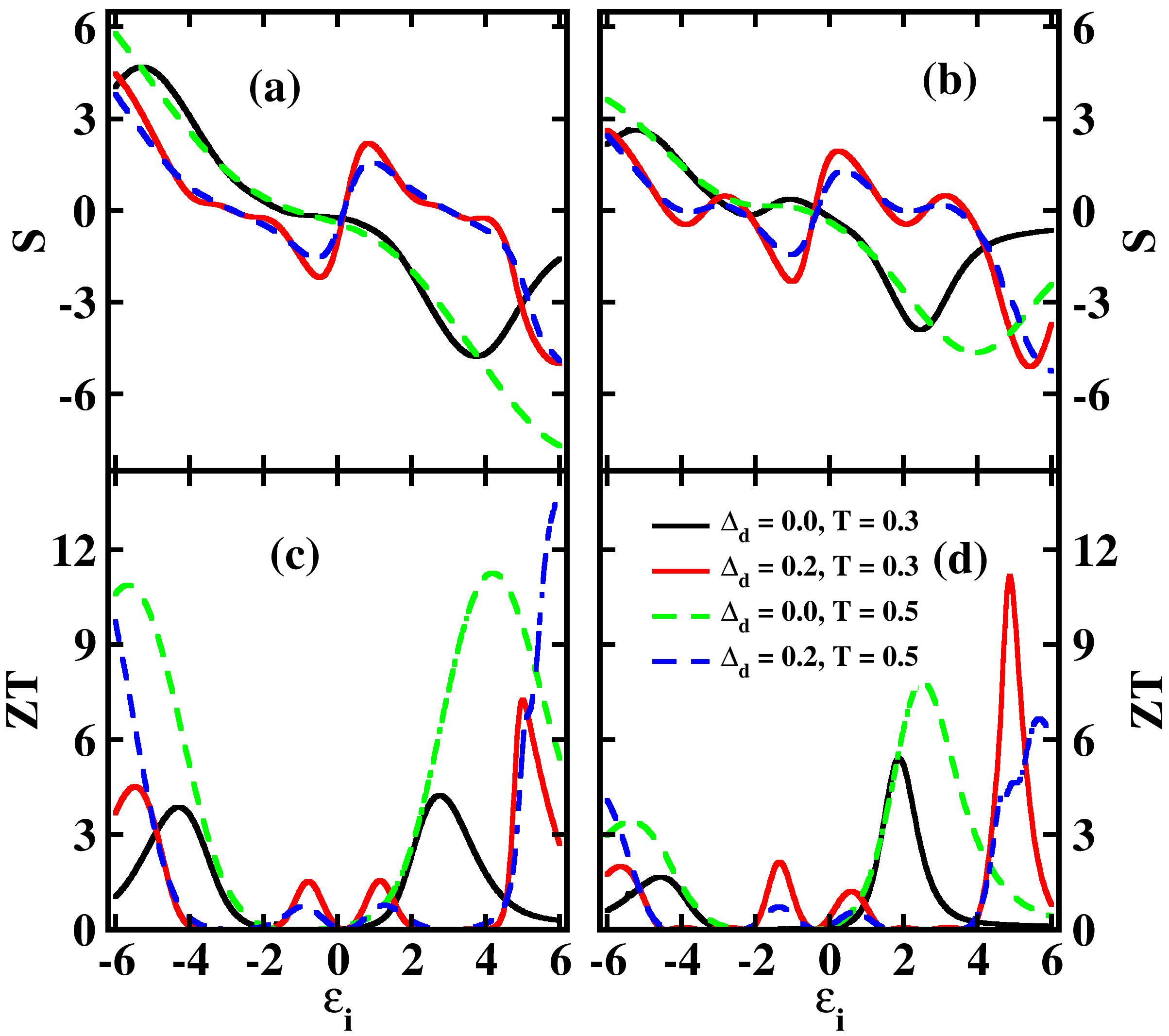}
	\caption{\label{s_zt}{Seebeck coefficient $S$ [(a),(b)] and figure of merit $ZT$ [(c),(d)] versus the normalized on-site energy $\varepsilon_i/\Gamma_0$ for both undriven and driven cases at two different temperatures. Panels (a),(c) show the pure ring and (b),(d) shows the ring--chord geometry.}}
\end{figure}

We now analyze the thermoelectric response of the system by focusing on the Seebeck coefficient $S$ and the dimensionless figure of merit $ZT$, which directly probe the energy selectivity of charge transport.
Fig.~\ref{s_zt} shows the  $S$ [(a) and (b)] and the $ZT$ [(c) and (d)] as functions of the $\varepsilon_i$ with variation of $\Delta_d$ and $T$. Panels (a) and (c) correspond to the pure ring geometry, while panels (b) and (d) represent the ring--chord geometry.

For the pure ring geometry, the Seebeck coefficient varies smoothly with $\varepsilon_i$ and undergoes sign changes as the dominant transport channel shifts from electron-like to hole-like.
In the absence of microwave driving, this gradual evolution reflects the relatively weak particle--hole asymmetry of the transmission function.
Consequently, the resulting figure of merit remains modest, with $ZT$ reaching values of approximately 4 only near resonant level alignment, where both electrical and thermal conductances increase simultaneously.

Upon application of the microwave field, photon-assisted tunneling generates additional Floquet sidebands that enhance the energy dependence of the transmission.
This leads to pronounced oscillations in the Seebeck coefficient and produces several local maxima in $ZT$.
However, in the pure ring geometry these enhancements remain limited, as the increase in thermopower is accompanied by a comparable enhancement of electronic heat transport.
As a result, even under driving, the maximum $ZT$ remains moderate, typically not exceeding values of $ZT \sim 7$ within the explored parameter range.

Please note that the maximum ZT is evaluated by sweeping the chemical potential with in a normalized energy range of $\pm6\Gamma_0$. This broad window is selected to cover the full spectral weight of the driven multi-dot system. Beyond these bounds ($|\varepsilon_i| > 6\Gamma_0$), the electrical conductance $G$ and electronic thermal conductance $\kappa$ fully diminish to zero (see Fig. \ref{gv_ke}). Sweeping over this window ensures that all relevant quantum channels, most notably the destructive Fano interference minima that supress $\kappa$, and the outer microwave-driven Floquet sidebands are fully considered. This confirms that the exceptional peak values of ZT are genuine physical manifestations of the Floquet-Fano energy-filtering mechanism rather than numerical artifacts of a truncated integration domain.  

A qualitatively distinct behavior emerges in the ring--chord geometry.
Even in the absence of microwave driving, the Seebeck coefficient exhibits sharper variations and significantly larger amplitudes than in the pure ring.
This enhancement originates from geometry-induced Fano interference, which introduces asymmetric line shapes and sharp antiresonances in the transmission spectrum.
Near these antiresonances, $S$ is strongly enhanced due to the steep energy dependence of the transmission around the Fermi level. The corresponding figure of merit shows pronounced peaks already in the undriven case, with $ZT$ reaching values around $ZT \approx 5$ near $\varepsilon_i \simeq 2$ (Fig. \ref{s_zt}(d)), where destructive interference suppresses thermal transport more efficiently than electrical conduction.

When microwave driving is applied, this effect is continueously amplified.
Photon-assisted processes dynamically reshape the Fano resonances, creating multiple narrow transmission zeros and steep slopes in the energy window relevant for transport. As a consequence, exceptionally large values of the thermoelectric efficiency are obtained.
In particular, for the driven ring--chord geometry, the figure of merit exhibits a sharp maximum exceeding $ZT \approx 11$ around $\varepsilon_i \simeq 5$.
This large enhancement arises from the simultaneous occurrence of a large thermopower and a strong suppression of the electronic thermal conductance.
The Floquet-induced sidebands act as an additional energy filter, effectively narrowing the transport window and enhancing particle--hole asymmetry without proportionally increasing heat flow.

The close correspondence between the transmission characteristics in Fig.~\ref{Transmission} and the thermoelectric response in Fig.~\ref{s_zt} highlights the physical origin of this behavior.
The largest values of $S$ and $ZT$ occur when the gate-controlled on-site energy aligns the Fermi level with Fano-suppressed regions of the transmission spectrum.
In these regimes, quantum interference selectively blocks high-energy carriers while preserving a strong energy derivative of the transmission, which is optimal for thermoelectric conversion.

The high $ZT$ values reported here are consistent with earlier recent theoretical studies predicting large thermoelectric efficiency in quantum-dot systems with sharp resonances, interference effects, or strong energy filtering~\cite{trocha2025spin,kuo2010thermoelectric,khedri2017influence}.
Importantly, the present enhancement is achieved purely through coherent electronic effects and Floquet engineering, without relying on phonon suppression mechanisms.
This demonstrates that the interplay between geometry-induced Fano interference and microwave driving provides a powerful and experimentally feasible route for optimizing thermoelectric performance in multi-quantum-dot nanostructures.

\subsection{Interplay of Thermal Broadening and Energy Filtering in ZT}

The temperature dependence of the thermopower $S$ and the figure of merit $ZT$ (in Fig.~\ref{s_zt}) reveals a contrasting yet inter-connected behavior in both ring and ring--chord geometries, with an additional sensitivity to the detuning parameter $\Delta_d$. With increasing temperature ($T=0.3 \to 0.5$), $S$ exhibits a reduction in peak amplitude and smoother oscillations due to thermal averaging of sharp energy-dependent features. This suppression is more pronounced for $\Delta_d=0.2$ (red and blue curves), particularly in the ring--chord configuration, where interference effects are stronger. In contrast, the maximum value of $ZT$ generally shows an enhancement with temperature, increasing from moderate values ($ZT \sim 4$ at $T=0.3$) to higher values reaching $ZT \sim 11$ at $T=0.5$ in the ring geometry. For $\Delta_d=0.2$, although high $ZT$ peaks (exceeding $ZT \sim 12$) are still observed, their positions and magnitudes become more sensitive to temperature due to stronger smearing of asymmetric transmission features. This behavior can be understood from the relation $ZT = S^2 G T / \kappa$: while increasing $T$ and $\kappa$ enhances carrier participation, the simultaneous suppression of $S$, especially for $\Delta_d=0.2$, can locally weaken $ZT$. Thus, the pure ring geometry exhibits a more robust enhancement of $ZT$, whereas the ring--chord system with finite detuning shows stronger temperature-induced degradation of interference-driven thermoelectric performance.

\begin{figure}[h!]
	\centering
	\includegraphics[width=1.0\columnwidth]{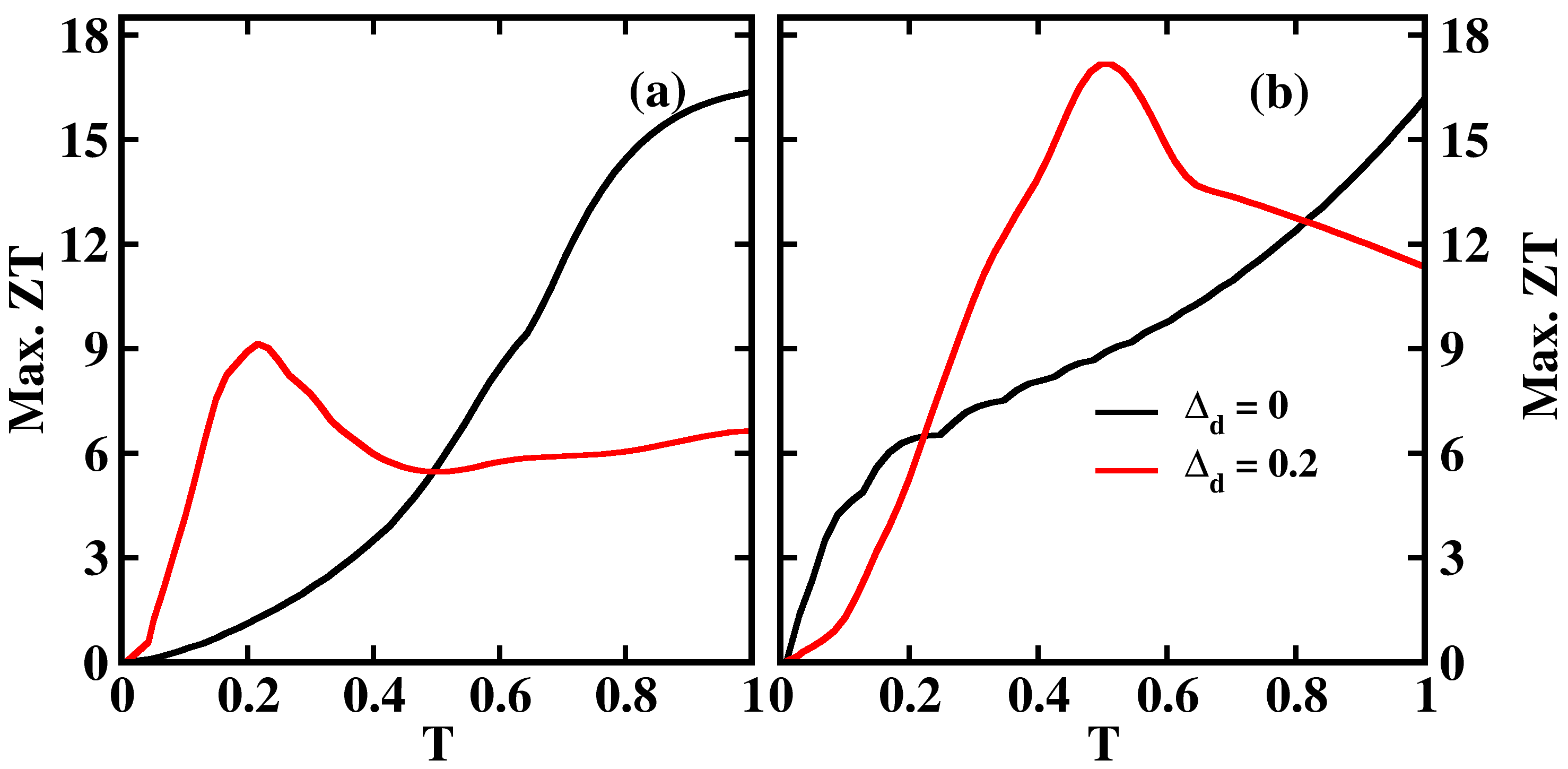}
	\caption{\label{max_zt_t}Temperature dependence of the maximum thermoelectric figure of merit for the (a) pure ring and (b) ring--chord geometries in the undriven ($\Delta_d=0$) and microwave-driven ($\Delta_d=0.2$) regimes.}
\end{figure}

To further clarify the influence of temperature on the thermoelectric efficiency, we next examine the evolution of the maximum thermoelectric figure of merit (Max.$\,ZT$) as a function of temperature for different quantum-dot geometries and detuning strengths.
The temperature dependence of the Max.$\,ZT$ for different quantum-dot geometries and detuning strengths is shown in Fig.~\ref{max_zt_t}. Panels (a) and (b) correspond to the four-site pure ring and four-site ring--chord systems, respectively.
The observed behavior can be understood more rigorously in terms of the energy-dependent transmission function and the associated transport integrals defined in Eq.~(\ref{eq:Ln}). Within the Landauer--B\"uttiker formalism, the thermoelectric transport coefficients are expressed through the moments of the transmission function.

For the quasi symmetric case ($\Delta_d = 0$), the transmission function remains nearly symmetric around the chemical potential. At low temperatures, the derivative $-\partial f/\partial E$ is sharply peaked at $E=\mu$, and only a narrow energy window contributes to the transport integrals in Eq.~(\ref{eq:Ln}). As temperature increases, this window broadens, allowing higher-order moments to contribute significantly. In this regime, the Seebeck coefficient [Eq.~(\ref{eq:Ssigma})] can be approximated using the Mott relation \cite{jonson1980mott},
	\begin{equation}
	S \approx -\frac{\pi^2 k_B^2 T}{3e}\left.\frac{d \ln \mathcal{T}(E)}{dE}\right|_{E=\mu},
	\end{equation}
	which shows that $S$ increases approximately linearly with temperature provided that the energy derivative of the transmission remains finite. Since the quasi-symmetric configuration maintains a smooth and nearly energy-symmetric transmission profile, the Seebeck coefficient increases gradually with temperature due to thermal broadening. At the same time, the absence of strong energy filtering prevents significant cancellation between electron-like and hole-like contributions in $L_1$ [Eq.~(\ref{eq:Ln})]. As a result, although the thermal conductance $\kappa$ [Eq.~(\ref{eq:kappa})] also increases with temperature, the combined growth of $S^2 G T$ dominates, leading to an overall monotonic increase of $ZT$ [Eq.~(\ref{eq:ZT})] in the this case.

In contrast, for finite detuning ($\Delta_d = 0.2$), the transmission function becomes strongly asymmetric due to the shift of energy levels. This enhances the energy filtering effect, producing a large slope $d\mathcal{T}/dE$ near the chemical potential and thus a strong increase in the Seebeck coefficient [Eq.~(\ref{eq:Ssigma})] at intermediate temperatures, consistent with the Mott relation. Consequently, $ZT$ [Eq.~(\ref{eq:ZT})] reaches a maximum when the thermal broadening optimally overlaps with the asymmetric transmission peak. However, at higher temperatures, the broadening of $-\partial f/\partial E$ in Eq.~(\ref{eq:Ln}) becomes too large, and contributions from both electron-like ($E>\mu$) and hole-like ($E<\mu$) excitations start to cancel in $L_1$. As a result, the Seebeck coefficient decreases, while the thermal conductance $\kappa$ [Eq.~(\ref{eq:kappa})], governed by $L_2$, continues to increase. This leads to a reduction in $ZT$ at higher temperatures.

The effect is further amplified in the ring--chord geometry, where Fano interference introduces sharp resonances and antiresonances in $\mathcal{T}(E)$. These features enhance the energy derivative $d\ln\mathcal{T}/dE$, thereby significantly increasing $L_1$ in Eq.~(\ref{eq:Ln}) and the Seebeck coefficient [Eq.~(\ref{eq:Ssigma})] at optimal temperatures. As a result, the peak value of Max.$\,ZT$ is higher and more pronounced compared to the pure ring geometry.

Overall, the temperature dependence of $ZT$ [Eq.~(\ref{eq:ZT})] is governed by the competition between the growth of the Seebeck coefficient (controlled by $L_1$ in Eq.~(\ref{eq:Ln})) and the increase of thermal conductance (controlled by $L_2$ in Eq.~(\ref{eq:Ln})).While $\Delta_d = 0$ favors a continuous increase due to gradual thermal activation, finite detuning ($\Delta_d \neq 0$) produces an optimal temperature where energy filtering is maximized.

\begin{figure}[h!]
	\centering
	\includegraphics[width=1.0\columnwidth]{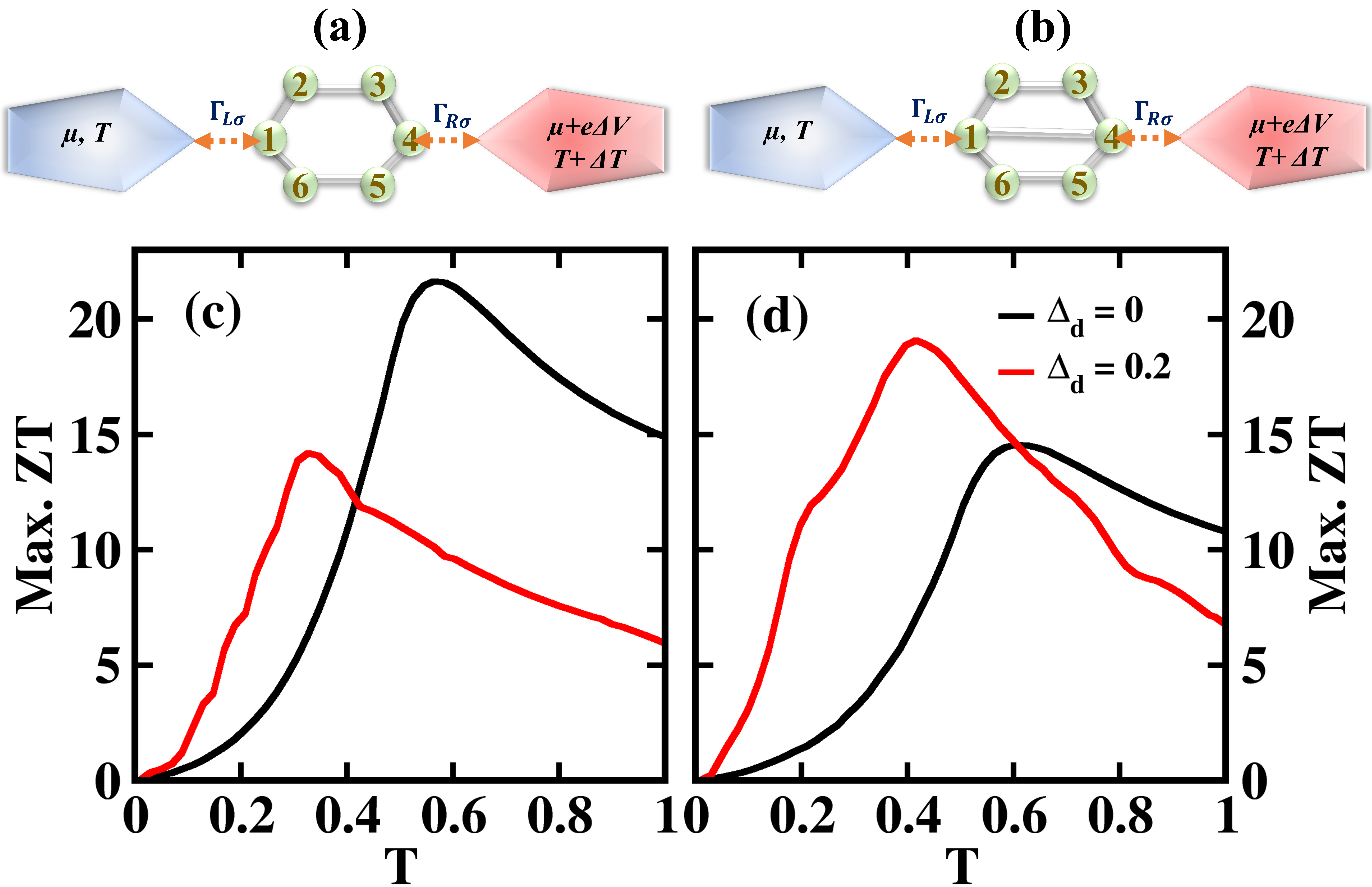}
	\caption{\label{six_max_zt_t}Schematic illustration of the six--quantum-dot nanostructures for the (a) ring-only and (b) ring--chord geometries. Panels (c) and (d) show the temperature dependence of the maximum thermoelectric figure of merit (Max.$,ZT$) for the pure ring and ring--chord systems, respectively, in both the undriven ($\Delta_d=0$) and microwave-driven ($\Delta_d=0.2$) regimes.}
\end{figure}

Although the main focus of the present work is the four-site ring and ring--chord structures, it is also important to examine whether the interference-assisted thermoelectric enhancement persists upon increasing the system size. To address this, we additionally investigate the corresponding six-site ring and ring--chord geometries shown in Fig.~\ref{six_max_zt_t}. This comparison provides insight into how enlarging the coherent transport network and increasing the number of available transport channels influence the thermoelectric response.

The schematic representations of the six-site pure ring and ring--chord geometries considered in this analysis are shown in Figs.~\ref{six_max_zt_t}(a) and \ref{six_max_zt_t}(b), respectively, while the corresponding temperature dependence of the maximum thermoelectric figure of merit is presented in Figs.~\ref{six_max_zt_t}(c) and \ref{six_max_zt_t}(d). Compared to the four-site geometries, the six-site systems exhibit a significantly enhanced thermoelectric response due to the larger number of available quantum states and coherent transport pathways.

In the six-site pure ring geometry [Fig.~\ref{six_max_zt_t}(c)], the undriven case $(\Delta_d = 0)$ shows a strong increase of Max.$ZT$ with temperature, reaching a peak value exceeding $20$ around $T \sim 0.55$, followed by a gradual decrease at higher temperatures. This behavior indicates that moderate thermal broadening enhances the contribution of resonant transport channels, thereby improving the balance between electrical conductance and thermopower. At higher temperatures, however, excessive thermal averaging weakens the energy selectivity of the transmission spectrum, leading to a reduction in $ZT$. In contrast, for the microwave-driven case $(\Delta_d = 0.2)$, Max.$ZT$ reaches a smaller peak value of about $14$ near $T \sim 0.3$ and then decreases continuously with temperature. The microwave field introduces additional photon-assisted transport channels that redistribute the spectral weight over a wider energy range, which reduces the sharpness of the resonant features and weakens the thermoelectric efficiency at higher temperatures.

A distinctly different temperature dependence appears in the six-site ring--chord geometry shown in Fig.~\ref{six_max_zt_t}(d), where the coexistence of multiple coherent transport pathways and Fano-type interference strongly modifies the thermoelectric response. For the undriven case $(\Delta_d = 0)$, Max.$ZT$ gradually increases with temperature and reaches a broad maximum of nearly $15$ around $T \sim 0.6$, after which it slowly decreases. The broader peak reflects the presence of multiple interference-induced resonances contributing over a wider temperature range. When microwave driving is introduced $(\Delta_d = 0.2)$, the thermoelectric performance is significantly enhanced at lower and intermediate temperatures, with Max.$ZT$ approaching $19$ near $T \sim 0.4$. This enhancement originates from the combined action of Floquet sidebands and Fano interference, which produces highly asymmetric transmission resonances and stronger energy filtering. At higher temperatures, thermal smearing progressively suppresses these sharp interference features, leading to a gradual decline in $ZT$. Nevertheless, the six-site ring--chord geometry maintains comparatively large thermoelectric efficiency compared to four-site geometry over a broad temperature range, demonstrating that increasing the system size further strengthens interference-assisted thermoelectric transport \cite{zimbovskaya2022large,sridhar2026coherent}.

\subsection{Efficiency at maximum power: interplay of geometry and microwave driving}

\begin{figure}[h!]
	\centering
	\includegraphics[width=1.0\columnwidth]{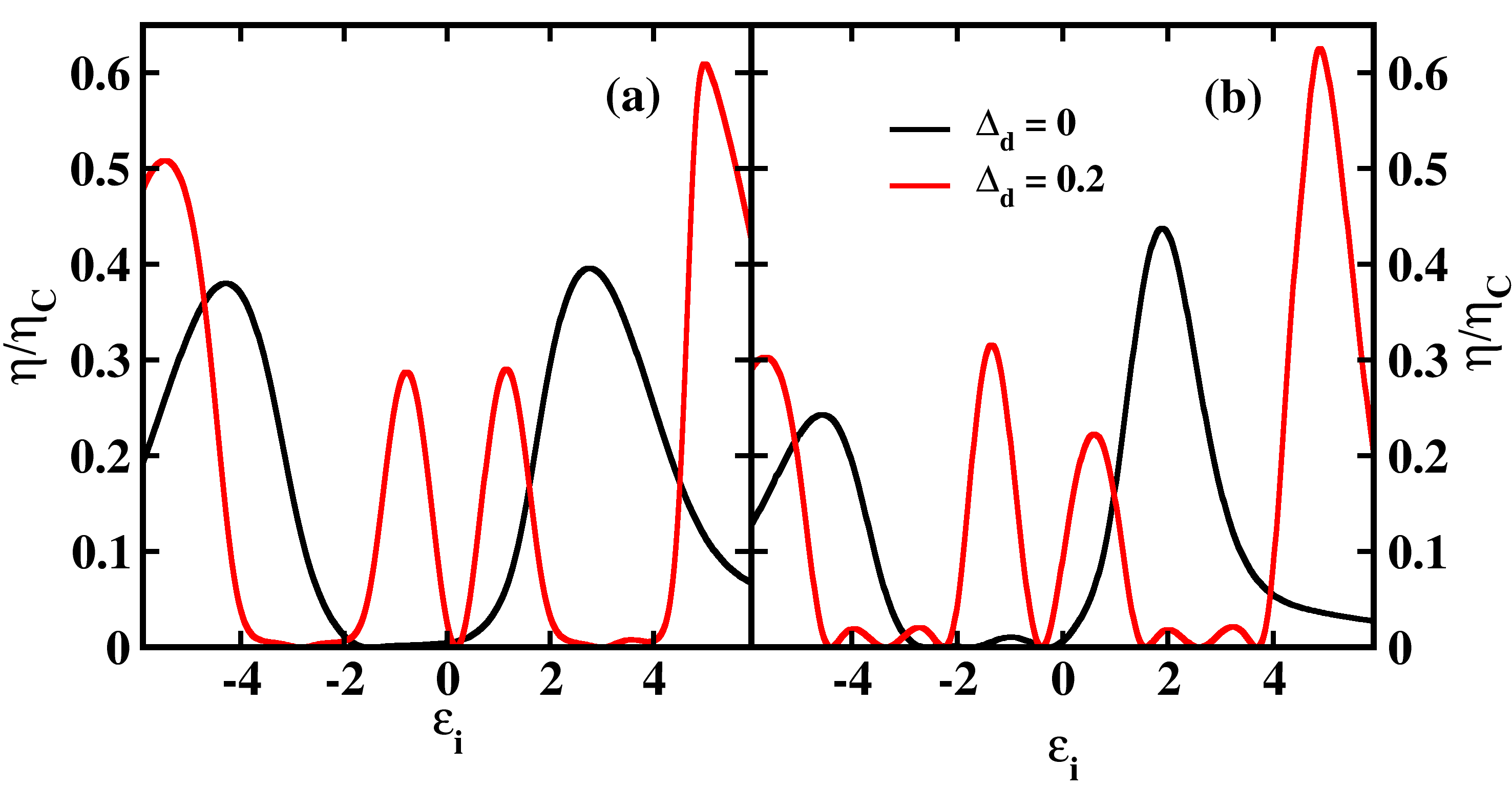}
	\caption{\label{efficiency}Efficiency at maximum power $\eta/\eta_C$ versus $\varepsilon_i$ for (a) pure ring and (b) ring--chord geometries, shown for $\Delta_d=0$ and $0.2$.}
\end{figure}

We now examine the efficiency at maximum power, normalized to the Carnot efficiency, $\eta/\eta_C$, as a function of the on-site energy $\varepsilon_i$.
This quantity provides a meaningful measure of thermoelectric performance beyond linear-response transport coefficients, as it directly addresses the practical exchange between convesion efficiency and power output.
The normalized efficiency, $\eta/\eta_C$, for the pure ring geometry and the ring--chord geometry is illustrated in Fig.~\ref{efficiency} (a) and (b), respectively.

For the pure ring configuration (Fig.
\ref{efficiency} (a)), the efficiency at maximum power exhibits relatively broad and smooth maxima as a function of $\varepsilon_i$ in the absence of microwave driving.
These features originate from resonant transport through extended ring eigenstates, where a compromise between electrical conductance, thermopower, and electronic thermal conductance yields finite efficiency.
When microwave driving is introduced, the efficiency profile undergoes substantial modification.
Photon-assisted tunneling generates additional Floquet sidebands that redistribute spectral weight, leading to a fragmentation of the efficiency peaks.
As a consequence, the driven efficiency displays multiple narrower maxima interspersed with regions of reduced efficiency, reflecting the competition between different photon-assisted transport channels.

A markedly different behavior is observed in the ring--chord geometry (Fig. \ref{efficiency} (b)).
Even without microwave driving, the efficiency shows sharper and more strongly modulated peaks, reaching values of $\eta/\eta_C \approx 0.42$.
This enhancement arises from geometry-induced Fano interference between the discrete chord-assisted transport path and the continuum of ring-mediated states.
Such interference selectively suppresses electronic thermal conductance while preserving sizable thermopower, thereby improving the efficiency at maximum power for specific gate-voltage configurations.

The application of microwave driving to the ring--chord system further amplifies this effect.
As shown by the red curve in the right panel, the efficiency can reach values as high as $\eta/\eta_C \approx 0.62$, exceeding those obtained in both the undriven ring--chord case and the microwave-driven pure ring.
This pronounced enhancement results from the combined action of Floquet sidebands and Fano interference, which dynamically reshape the transmission spectrum.
Photon-assisted processes enable a simultaneous optimization of electrical conductance and thermopower, while interference effects continue to suppress thermal transport.

Overall, results from Fig.~\ref{efficiency} highlights the complementary roles of system geometry and time-periodic driving in controlling thermoelectric efficiency.
While microwave driving alone reshapes the efficiency landscape of the pure ring, the inclusion of a chord introduces strong Fano interference that, when combined with Floquet engineering, leads to substantial gains in efficiency at maximum power.
These findings underscore the potential of ring--chord quantum-dot architectures as highly tunable and efficient platforms for nanoscale thermoelectric applications.

 \subsection{Power--Efficiency Trade-Off via Quantum Interference}
 
\begin{figure}[h!]
	\centering
	\includegraphics[width=1.0\columnwidth]{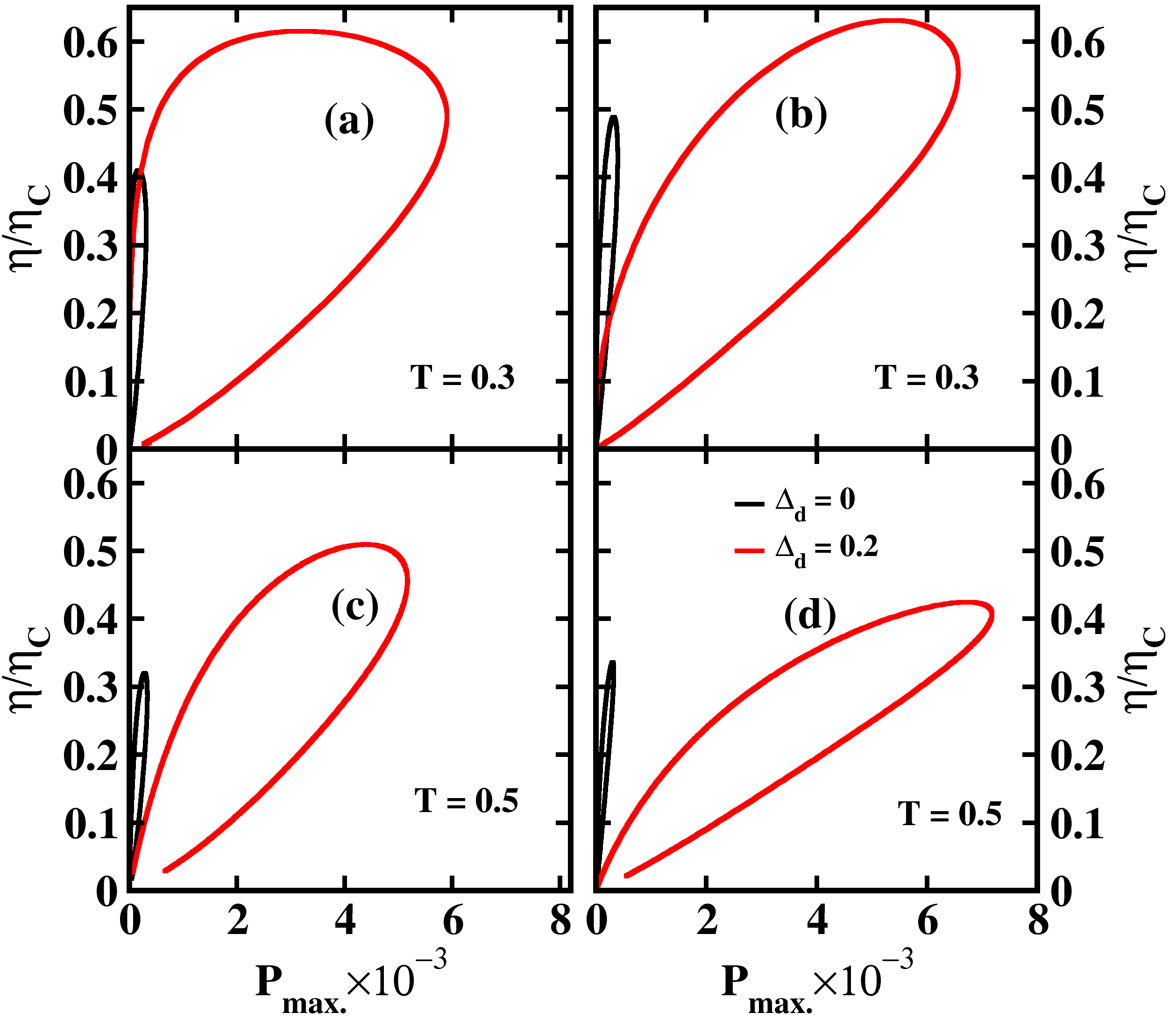}
	\caption{\label{pmax}Normalized efficiency $\eta/\eta_C$ as a function of the maximum output power $P_{\mathrm{max}}$ for the four-quantum-dot thermoelectric system. Panels (a) and (b) correspond to the ring-only and ring--chord geometries at $T=0.3$, while panels (c) and (d) represent the corresponding geometries at $T=0.5$. Black and red curves denote $\Delta_d=0.0$ and $\Delta_d=0.2$, respectively.}
\end{figure}

To quantify the overall performance and practical viability of the proposed heat engines it is essential to examine the simultaneous optimization of power and efficiency. For this the normalized efficiency $\eta/\eta_C$ as a function of the maximum output power $P_{\mathrm{max}}$ for two different configurations of the four-quantum-dot thermoelectric system are illustrated in	Fig.~\ref{pmax}. The black and red curves correspond to $\Delta_d = 0.0$ and $\Delta_d = 0.2$, respectively. Panels (a) and (b) correspond to the ring-only and ring--chord geometries at temperature $T=0.3$, while panels (c) and (d) represent the same geometries at the higher temperature $T=0.5$. A pronounced enhancement in both the achievable efficiency and the output power is observed upon introducing finite $\Delta_d$, particularly in the ring--chord configuration.

For $\Delta_d = 0.0$, the efficiency--power characteristics remain confined to a narrow region near the origin, indicating relatively weak thermoelectric performance. The maximum normalized efficiency reaches only about $0.4\eta_C$, while the corresponding output power remains very small. This behavior originates from the nearly symmetric transmission spectrum of the system, where electron and hole contributions around the chemical potential partially cancel each other, thereby suppressing the thermopower and limiting the power factor.
In contrast, when $\Delta_d = 0.2$, the efficiency--power loop expands significantly in both geometries. The normalized efficiency approaches nearly $0.62\,\eta_C$, while the maximum output power increases by more than an order of magnitude. The enhancement is more pronounced in the ring--chord configuration shown in Fig.~\ref{pmax}(b), where the loop becomes broader and extends to higher values of $P_{\mathrm{max}}$. This demonstrates that the additional hopping pathway generates strong quantum interference effects that simultaneously improve the efficiency and the electrical power generation.

At the higher temperature $T=0.5$, shown in Figs.~\ref{pmax}(c) and \ref{pmax}(d), the efficiency--power loops become comparatively compressed along the efficiency axis, although the accessible power output remains finite over a broader range. In the ring-only geometry [Fig.~\ref{pmax}(c)], the maximum normalized efficiency decreases compared to the low-temperature case due to thermal broadening of the transmission spectrum. A similar reduction is observed in the ring--chord geometry [Fig.~\ref{pmax}(d)], where the peak efficiency is lower than that in Fig.~\ref{pmax}(b), although the system still preserves comparatively larger power output for $\Delta_d=0.2$. The reduction in efficiency at higher temperature originates from enhanced electron-hole excitations around the Fermi energy, which weaken the transmission asymmetry responsible for large thermopower. Nevertheless, the ring--chord configuration continues to exhibit superior thermoelectric performance compared to the simple ring geometry, demonstrating the robustness of quantum-interference-assisted transport even in the presence of stronger thermal fluctuations.

The observed behavior can be understood within the Landauer transport framework, where the output power is determined by the competition between the Seebeck coefficient and the electrical conductance. In the ring--chord geometry, electron transport occurs through multiple coherent pathways, and the transmission probability can be expressed as
 \begin{equation}
 	\mathcal{T}(E) \sim \left|A_1(E)+A_2(E)\right|^2,
 \end{equation}
 where $A_1(E)$ and $A_2(E)$ denote the amplitudes associated with different transport channels. The interference between these channels produces sharp resonant and antiresonant features in the transmission spectrum, resulting in a strong energy asymmetry around the Fermi energy. Such asymmetry enhances the thermopower while maintaining finite conductance, thereby increasing the power factor and improving the thermoelectric conversion efficiency. At higher temperatures, thermal broadening partially smoothens these resonant and antiresonant features, leading to the reduced peak efficiencies observed in Figs.~\ref{pmax}(c) and \ref{pmax}(d).
 
 Another important feature visible in Fig.~\ref{pmax} is the enlarged enclosed area of the red loops for $\Delta_d = 0.2$, particularly at the lower temperature $T=0.3$. This indicates an improved balance between efficiency and power output. In conventional thermoelectric systems, achieving high efficiency generally requires operation near the open-circuit condition where the output power becomes negligible. Here, however, the interference-induced asymmetry allows the system to sustain comparatively high efficiency even at finite power output, indicating a substantial reduction of the usual efficiency--power trade-off.
These results possess strong experimental relevance for nanoscale thermoelectric devices. Similar interference-driven enhancement mechanisms have been experimentally observed in Aharonov--Bohm interferometers, coupled quantum dots, molecular junctions, and low-dimensional semiconductor nanostructures, where coherent transport plays a dominant role \cite{josefsson2019optimal,kwon2024optimization,kuo2017large,sridhar2026coherent}. In practical realizations, the parameter $\Delta_d$ may correspond to tunable asymmetry induced through gate voltages, unequal dot--lead couplings, orbital detuning, strain engineering, or magnetic-field-controlled level splitting. For example, in quantum-dot heat engines, gate electrodes can dynamically tune the energy levels and interdot coupling strengths, thereby controlling the interference pattern and optimizing the thermoelectric response. Likewise, in molecular junctions, asymmetric molecule--electrode hybridization can generate transmission antiresonances similar to those predicted here.
 Therefore, the present results demonstrate that controlled quantum interference provides an effective and experimentally feasible route for enhancing both the efficiency and power generation capability of nanoscale thermoelectric heat engines.

\subsection{Spin-dependent thermoelectric transport}
\subsubsection{Spin-dependent electronic conductance}
\begin{figure}[h!]
	\centering
	\includegraphics[width=1.0\columnwidth]{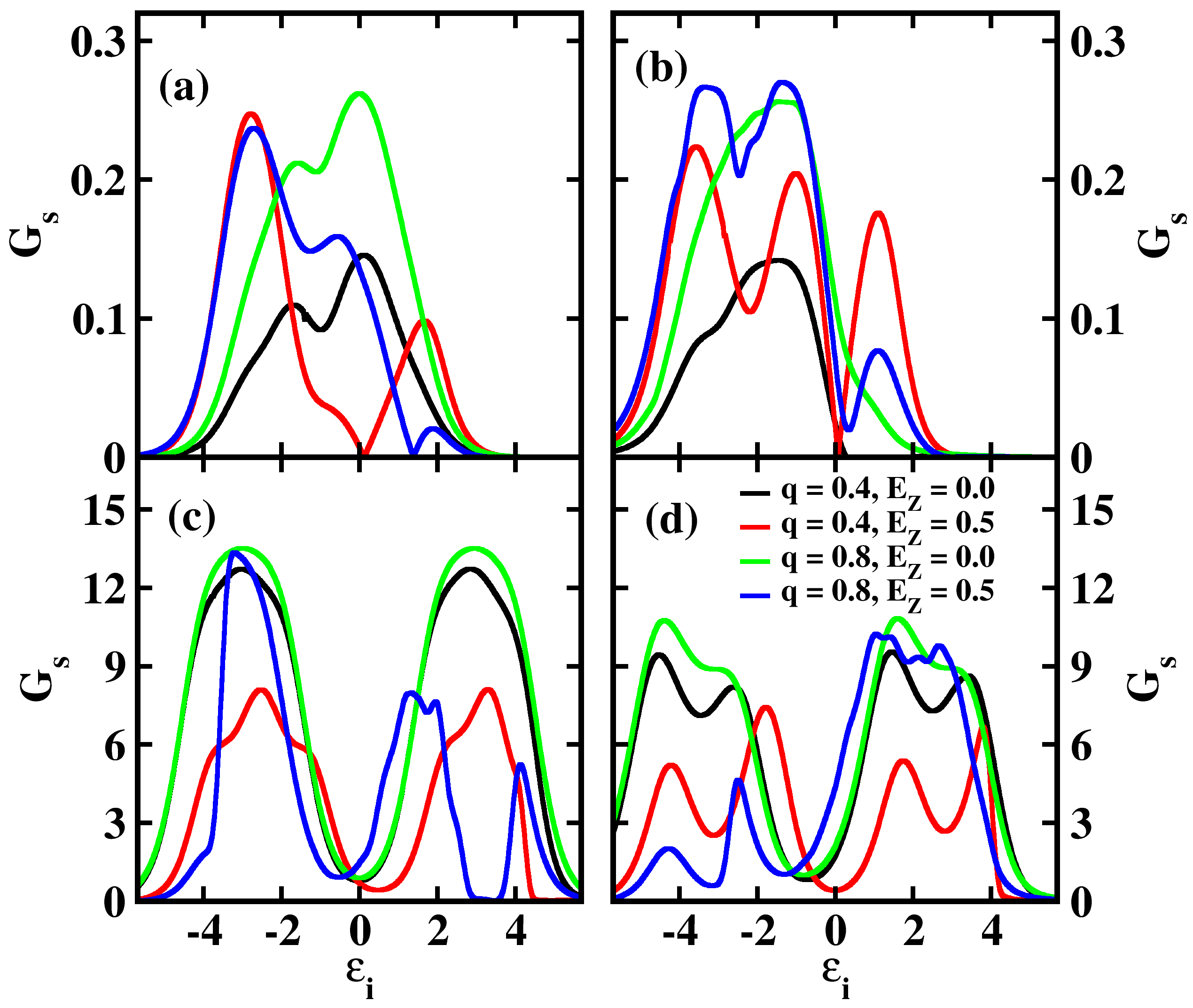}
	\caption{\label{Gs_all}Spin-dependent electronic conductance $G_s$ as a function of the on-site energy $\varepsilon_i$ for the pure ring geometry (left panels) and the ring--chord geometry (right panels). The upper panels correspond to the undriven case ($\Delta_d = 0$), while the lower panels show the microwave-driven regime with $\Delta_d = 0.2$.}
\end{figure}

To explore how spin polarization and quantum interference influence the thermoelectric response of the system, we now investigate the spin-dependent thermoelectric transport properties of the four--quantum dots systems. We first present the spin-dependent electronic conductance $G_s$ as a function of the on-site energy $\varepsilon_i$ for the pure ring geometry ((a) and (c)) and the ring--chord geometry ((b) and (d)) in Fig.~\ref{Gs_all}.
The upper panels correspond to the undriven case ($\Delta_d = 0$), while the lower panels show the microwave-driven regime with $\Delta_d = 0.2$.
Different curves represent various values of the lead polarization $q$ and the Zeeman splitting $E_z$, as indicated.

In the absence of a magnetic field, the spin-up and spin-down channels are degenerate, resulting in a relatively simple conductance spectrum characterized by a limited number of resonant peaks.
Applying a finite magnetic field lifts this degeneracy through Zeeman splitting, leading to a separation of spin-resolved transport channels.
Consequently, the number of conductance peaks increases and their positions shift in energy according to $E_z$, reflecting spin-dependent resonant tunneling processes.

For the pure ring geometry (Fig. \ref{Gs_all} (a)) in the undriven case, the spin-dependent conductance exhibits relatively broad resonances.
Peaks located at $\varepsilon_i \simeq \pm E_z$ correspond to the tunneling of a spin-up or spin-down electron into an otherwise empty structure.
Additional resonances at shifted energies $\varepsilon_i \simeq \pm E_z - U$ arise from transport through states where the system is already occupied by an electron of opposite spin. The magnitude of these resonances is further modulated by the lead polarization q. As q increases from 0.4 to 0.8, we observe a significant enhancement in the conductance amplitudes (green and blue curves). This indicates that higher lead polarization facilitates more efficient injection into the spin-split levels, amplifying the spin-filtering effect of the ring. Increasing the magnetic field not only shifts the peak positions but also enhances their amplitudes, indicating increased spin accumulation and stronger spin population imbalance within the quantum dot system.

When the microwave field is applied ($\Delta_d = 0.2$, Fig. \ref{Gs_all} (c)), photon-assisted tunneling opens additional spin-resolved transport channels.
Each resonant feature is accompanied by sidebands separated by integer multiples of the photon energy, giving rise to new conductance peaks at energies $\varepsilon_i \simeq \pm E_z - U \pm n\hbar\omega$, where $n$ is an integer.
This process substantially enriches the spin conductance spectrum and activates transport channels that are absent in the undriven regime. Notably, the microwave driving leads to a dramatic increase in the conductance magnitude, with peak values reaching $G_s \approx 14$, nearly two orders of magnitude higher than the static case. As in the undriven regime, the lead polarization q continues to act as a scaling factor; the green and blue curves (q=0.8) exhibit much larger amplitudes and sharper features compared to the lower polarization cases (q=0.4), underscoring the synergy between Floquet engineering and spin-polarized injection. 

The ring--chord geometry (right column) exhibits qualitatively distinct behavior due to geometry-induced Fano interference.
Even without microwave driving (Fig. \ref{Gs_all} (b)), sharp asymmetric features appear in $G_s$ as a result of quantum interference between different transport paths.
Destructive interference selectively suppresses one spin channel at specific energies, while the opposite spin channel remains conductive. This leads to strongly spin-polarized conductance peaks whose positions and amplitudes are highly sensitive to the magnetic field and lead polarization.

In the presence of microwave driving (Fig. \ref{Gs_all} (d)), the interplay between Floquet sidebands and Fano interference produces a dense pattern of sharp resonances and antiresonances in the spin conductance.
Photon-assisted processes significantly enhance the peak amplitudes of $G_s$ and introduce additional spin-resolved features, while interference effects continue to control the energy selectivity of transport.
Remarkably, the sharp dips (antiresonances) characteristic of the ring-chord geometry remain clearly visible even under strong driving. This suggests that the geometry-induced Fano filtering is robust against photon-induced broadening, maintaining its ability to suppress transport at specific energies.
As a result, the driven ring--chord system displays a much stronger and more tunable spin-dependent conductance than in the undriven case. By utilizing the microwave field to amplify the signal and the chord to provide sharp energy selectivity, the system acts as a high-performance, dynamically controlled spin filter. 

Overall, the spin electronic conductance is governed by the combined effects of magnetic field, lead polarization, system geometry, and microwave driving.
The magnetic field controls the energy separation and magnitude of spin-resolved resonances through spin accumulation, while the microwave field increases the number and strength of available transport channels.
The ring--chord geometry further increases these effects via Fano interference, making it a particularly effective platform for engineering spin-selective transport in nanoscale devices.

\subsubsection{Spin thermopower}

\begin{figure}[h!]
	\centering
	\includegraphics[width=1.0\columnwidth]{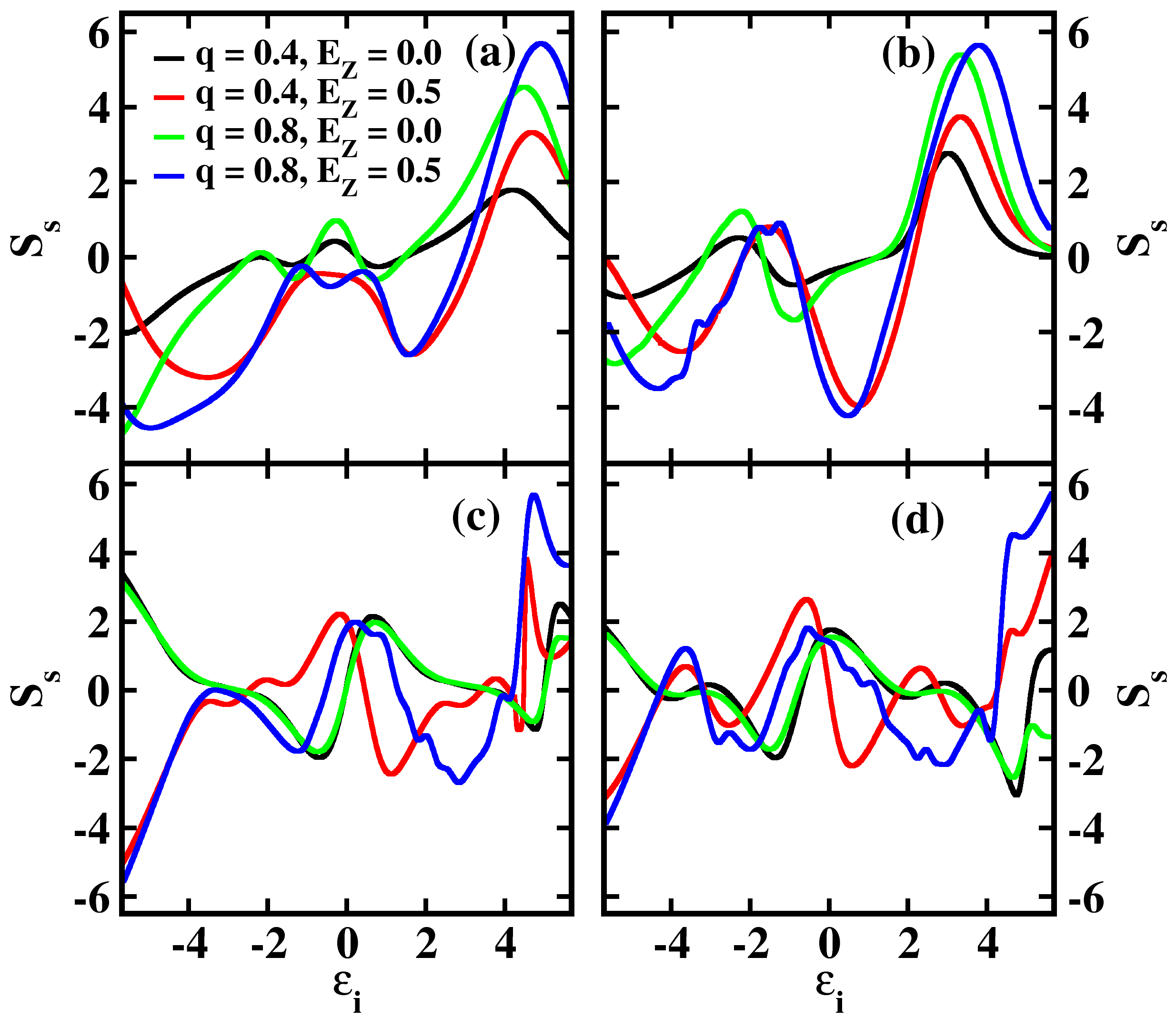}
	\caption{\label{Ss_all}Spin thermopower $S_s$ as a function of $\varepsilon_i$ for the pure ring geometry (left panels) and the ring--chord geometry (right panels). The upper panels correspond to the undriven case ($\Delta_d = 0$), while the lower panels show the microwave-driven regime with $\Delta_d = 0.2$.}
\end{figure}

To evaluate the spin-selective heat-to-electricity conversion capabilities of the system, we examine the thermoelectric response in the presence of magnetic and microwave field. The spin thermopower $S_s$ versus $\varepsilon_i$ is shown in Fig.~\ref{Ss_all}, where (a),(c) correspond to the pure ring and (b),(d) to the ring--chord geometry.

We first discuss the behavior of the spin thermopower in the absence of microwave driving.
For both geometries, $S_s$ exhibits sign changes as $\varepsilon_i$ is varied, reflecting transitions between spin-up–dominated and spin-down–dominated transport.
At energies close to resonance and near particle--hole symmetry, the contributions of opposite spin channels compensate each other, leading to a suppression of the spin thermopower.
Away from these points, spin-dependent asymmetry in the transmission becomes pronounced, resulting in finite values of $S_s$.
In the pure ring geometry (Fig. \ref{Ss_all} (a)), the spin thermopower varies smoothly with $\varepsilon_i$.
For zero magnetic field, $S_s$ remains relatively small even at finite lead polarization, indicating weak spin selectivity.
Applying a magnetic field lifts the spin degeneracy and shifts the spin-resolved resonances, which enhances the spin imbalance between transport channels.
As a result, the magnitude of $S_s$ increases noticeably, particularly for strongly polarized leads ($q = 0.8$), where $|S_s|$ reaches values of order $4$--$5$, signalling a high degree of spin-thermoelectric efficiency.

A qualitatively different response is observed in the ring--chord geometry (Fig. \ref{Ss_all} (b)).
Here, geometry-induced Fano interference introduces sharp energy-dependent asymmetry in the spin-resolved transmission.
These interference effects result in sharper oscillations and larger amplitudes of the spin thermopower compared to the pure ring case.
In the presence of both finite polarization and magnetic field, $S_s$ exhibits pronounced extrema, highlighting that the energy filtering is significantly enhanced by the interference of the chord architecture.

When the microwave field is applied, photon-assisted tunneling significantly modifies the spin thermopower.
Additional Floquet sidebands appear in the transmission spectrum, leading to more rapid oscillations of $S_s$ as a function of $\varepsilon_i$.
In the pure ring geometry (Fig. \ref{Ss_all} (c)), the microwave field increases the number of oscillations but slightly reduces the peak magnitude of $S_s$ due to the redistribution of spectral weight among multiple photon-assisted channels.

In contrast, the microwave-driven ring--chord system (Fig. \ref{Ss_all} (d)) shows a complex pattern of sharp oscillations with large amplitude.
The interplay between Floquet sidebands and Fano interference produces strong spin-dependent asymmetry over a broad energy range.
Although the microwave field tends to reduce the absolute value of $S_s$ by opening additional transport channels, the presence of a magnetic field counteracts this effect by enhancing spin splitting and spin accumulation.
As a result, significant values of the spin thermopower are preserved even in the driven regime.

Overall, these results demonstrate that the spin thermopower is highly sensitive to system geometry, magnetic field, lead polarization, and microwave driving.
While the pure ring geometry exhibits relatively smooth and moderate spin thermopower, the ring--chord configuration enables strong enhancement through interference effects.
Microwave driving further enriches the spin thermopower response by introducing photon-assisted processes, offering an additional tuning knob for spin caloritronic control.

\subsubsection{Spin thermoelectric figure of merit}
\begin{figure}[h!]
	\centering
	\includegraphics[width=1.0\columnwidth]{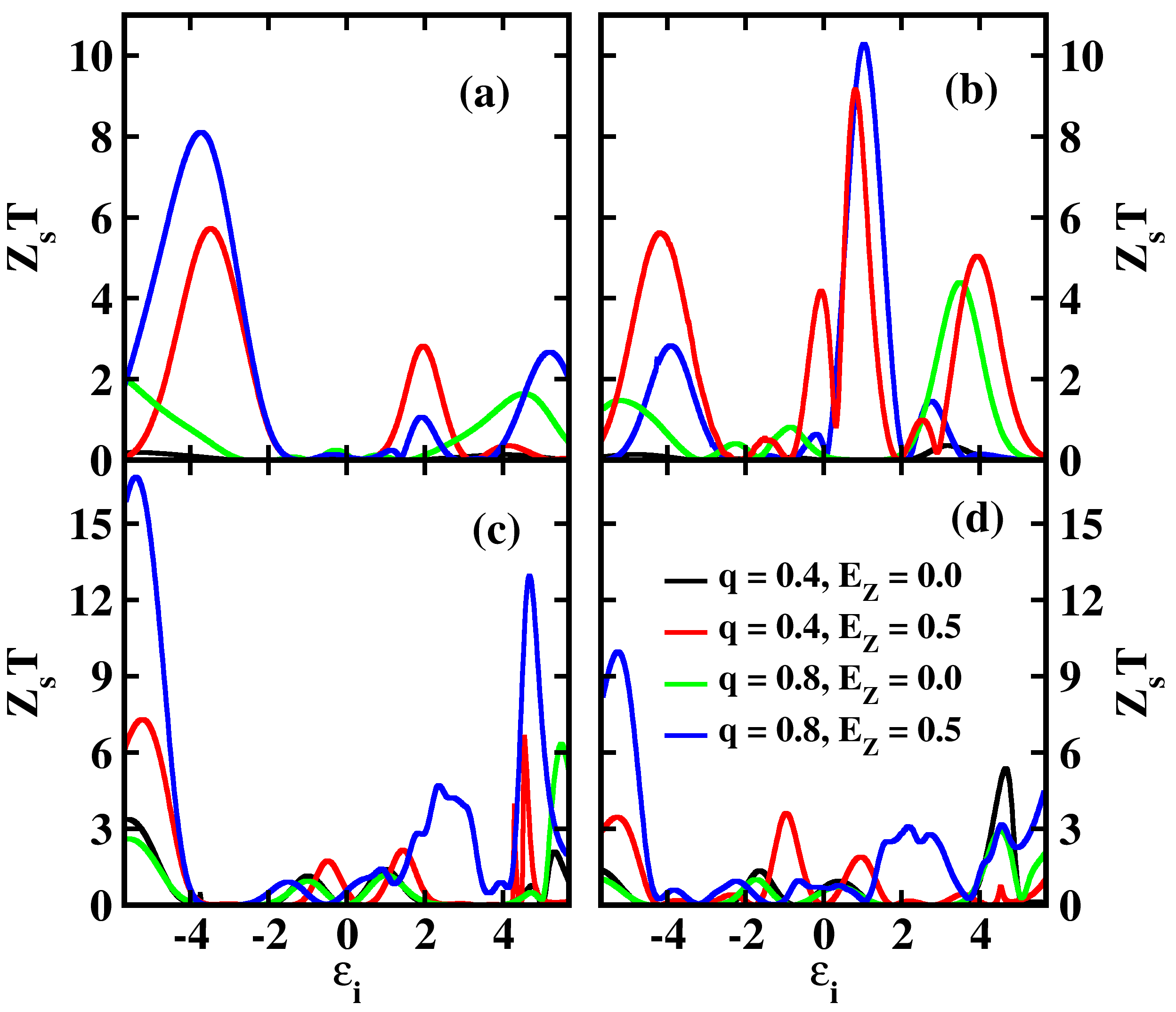}
	\caption{\label{zst_all_2D}Spin thermoelectric figure of merit $Z_{s}T$ as a function of $\varepsilon_i$ for the pure ring geometry (left panels) and the ring--chord geometry (right panels). The upper panels correspond to the undriven case ($\Delta_d = 0$), while the lower panels show the microwave-driven regime with $\Delta_d = 0.2$.}
\end{figure}

To quantify the overall efficiency of the spin-selective thermoelectric conversion, we calculate the spin-dependent dimensionless figure of merit $Z_sT$. This parameter accounts for the competition between the enhanced thermopower and the dissipative effects of thermal and electrical conductance.
The spin thermoelectric figure of merit $Z_sT$ as a function of the on-site energy $\varepsilon_i$ is presented in Fig.~\ref{zst_all_2D}, where panels (a) and (c) correspond to the pure ring geometry, while panels (b) and (d) represent the ring--chord geometry.

For the pure ring geometry, the spin figure of merit remains relatively small when the microwave field is absent.
As shown in Fig.~\ref{zst_all_2D}(a), the spin thermoelectric response remains weak in the absence of magnetic-field-induced spin splitting. For weakly polarized leads ($q = 0.4$) and zero magnetic field ($E_z = 0.0$), the spin figure of merit is nearly suppressed throughout the considered energy range. Increasing the lead polarization to $q = 0.8$ enhances the spin asymmetry in the transport channels, resulting in a moderate increase of the spin figure of merit up to $Z_sT \approx 2$. However, the overall enhancement remains limited in the absence of Zeeman splitting. A substantial increase in $Z_sT$ emerges when a finite magnetic field ($E_z = 0.5$) is introduced. In this regime, the interplay between Zeeman-induced spin splitting and lead polarization generates pronounced resonant features, yielding a maximum value of $Z_sT \approx 8$ around $\varepsilon_i \simeq -4$ for the highly polarized case ($q = 0.8$).
This enhancement reflects the increased spin imbalance in both the spin-dependent conductance and the spin Seebeck coefficient.

The application of microwave driving leads to a substantial enhancement of the spin thermoelectric efficiency in the pure ring system, as shown in the Fig. \ref{zst_all_2D} (c).
Photon-assisted tunneling opens additional spin-resolved transport channels, giving rise to multiple sharp resonances in $Z_sT$.
In this regime, the spin figure of merit reaches values as large as $Z_sT \approx 16$ for strongly polarized leads ($q = 0.8$) in the presence of a magnetic field ($E_z=0.5$).
These large values originate from the simultaneous enhancement of the spin thermopower and the suppression of the spin-dependent thermal conductance near photon-assisted resonance points, signalling a high efficient spin-caloritronic performance.

A qualitatively different behavior is observed in the ring--chord geometry due to geometry-induced Fano interference.
Even without microwave driving (Fig. \ref{zst_all_2D} (b)), sharp and high-amplitude peaks emerge in $Z_sT$.
For strongly polarized leads ($q = 0.8$) and finite Zeeman splitting ($E_z = 0.5$), the spin figure of merit reaches values close to $Z_sT \approx 10$ near $\varepsilon_i \simeq 1$, significantly exceeding those obtained in the pure ring geometry.
This enhancement results from interference-induced suppression of one spin channel, which strongly amplifies the spin Seebeck coefficient while reducing the spin thermal conductance.

When the microwave field is applied to the ring--chord system (Fig. \ref{zst_all_2D} (d)), the interplay between Floquet sidebands and Fano interference produces a dense pattern of sharp resonances in $Z_sT$.
Although the number of peaks increases considerably, the maximum value of the spin figure of merit is slightly reduced compared to the undriven case, with typical peak values around $Z_sT \sim 9$.
This minor reduction arises from the redistribution of spectral weight among multiple photon-assisted channels, which lowers the peak transmission probability and consequently reduces both the spin-dependent conductance and the spin Seebeck coefficient.
In contrast, the spin thermal conductance does not decrease proportionally, as it depends not only on transmission probabilities but also on the energy carried by the charge carriers.

Overall, the spin thermoelectric figure of merit exhibits trends similar to those of the conventional thermoelectric efficiency, but with a much stronger sensitivity to magnetic field and lead polarization.
For instance, in the undriven ring geometry (Fig. \ref{zst_all_2D} (a)), a pronounced peak occurs near $\varepsilon_i \simeq -4$.
At this energy, the spin figure of merit reaches $Z_sT \approx 8$ for $(q=0.8,\,E_z=0.5)$, while it is $Z_sT \approx 6$ for $(q=0.4,\,E_z=0.5)$, corresponding to an enhancement factor of $\approx 8/6 \simeq 1.3$ upon increasing the lead polarization.
At fixed polarization $q=0.8$, turning on the magnetic field from $E_z=0$ to $E_z=0.5$ increases the peak from $Z_sT \approx 2$ to $Z_sT \approx 8$, i.e., an amplification by a factor of $\approx 8/2 \simeq 4$. These results clearly highlight the dominant role of the magnetic field in controlling the spin thermoelectric response, and the large $Z_sT$ values obtained here are consistent with those reported in earlier theoretical studies\cite{hong2020large,trocha2012large}.
So, our results demonstrate that the spin thermoelectric efficiency can be efficiently tuned through the combined effects of system geometry, lead polarization, magnetic field, and microwave driving, with the ring--chord configuration offering a particularly promising platform for spin caloritronic applications.

\subsubsection{Optimization of the Maximum Spin Thermoelectric Figure of Merit}

\begin{figure}[h!]
	\centering
	\includegraphics[width=1.0\columnwidth]{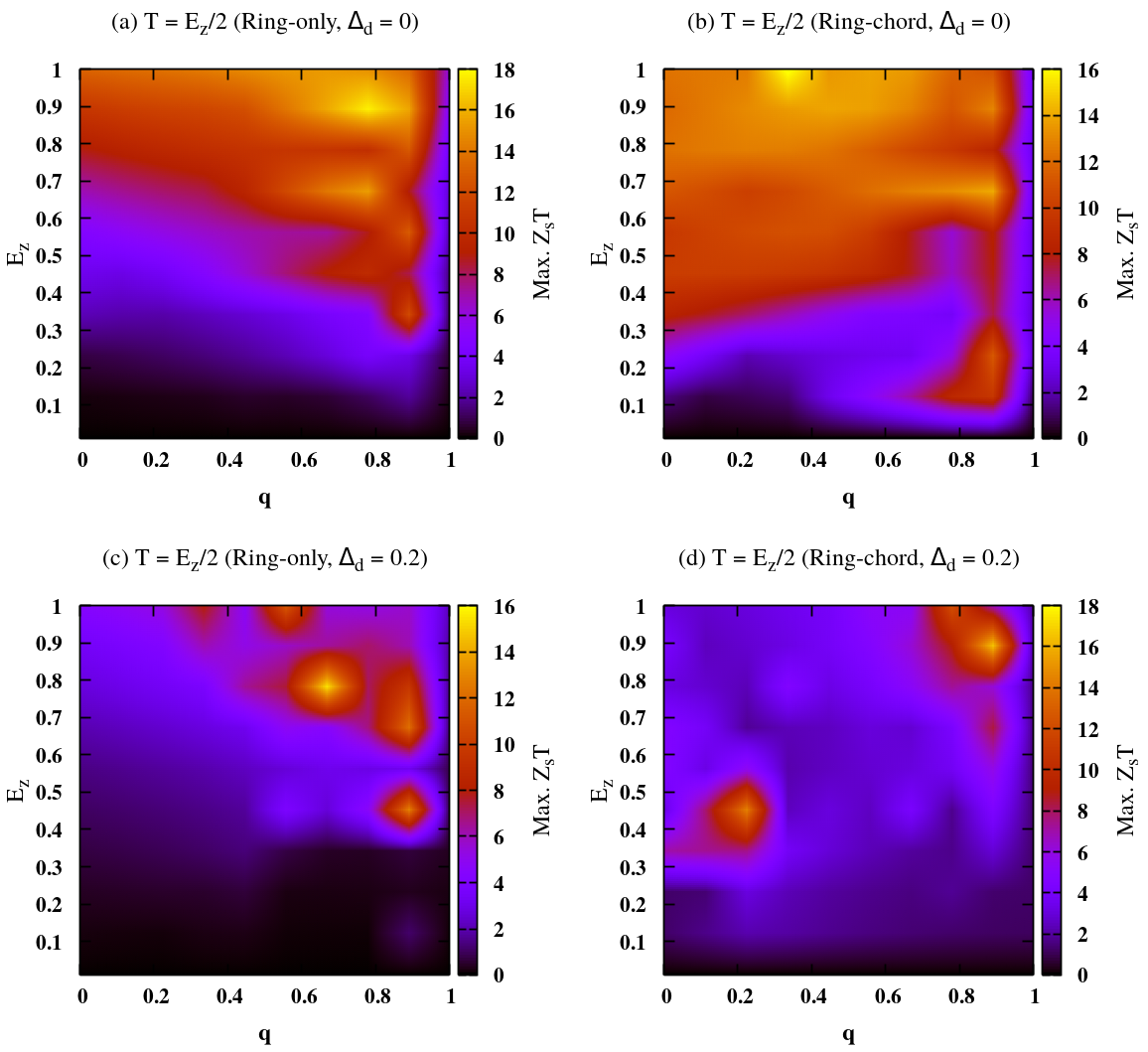}
	\caption{\label{ZsT_all}Maximum spin thermoelectric figure of merit Max.$\,Z_sT$ as functions of the lead polarization $q$ and Zeeman splitting $E_z$ at the temperature condition $T=E_z/2$. Panels (a) and (c) correspond to the pure ring geometry, while panels (b) and (d) represent the ring--chord geometry. The upper panels [(a) and (b)] are shown for $\Delta_d=0$, whereas the lower panels [(c) and (d)] correspond to finite detuning $\Delta_d=0.2$.}
\end{figure}

To further clarify the parameter regimes that maximize the spin thermoelectric response, we map the optimized spin figure of merit (Max.$,Z_sT$) in the two-dimensional $(q,E_z)$ parameter space for both the pure ring and ring--chord geometries. The maximum values are extracted by optimizing the on-site energy within the interval $-6 \leq \varepsilon_i \leq 6$, which fully covers all relevant transport resonances and interference-induced features. The contour plots of Max.$,Z_sT$ at the temperature condition $T=E_z/2$ for both the undriven and microwave-driven regimes is shown in Fig.~\ref{ZsT_all}.

For the pure ring geometry [Figs.~\ref{ZsT_all}(a) and (c)], the high-efficiency regions are relatively broad, indicating that the spin thermoelectric performance remains robust over an extended range of lead polarization and Zeeman splitting. In the absence of microwave driving [$\Delta_d=0$, Fig.~\ref{ZsT_all}(a)], Max.$Z_sT$ gradually increases with both $q$ and $E_z$, reaching values close to $18$ in the strong spin-polarization regime. This behavior reflects the progressive enhancement of spin-selective transport caused by the combined action of ferromagnetic leads and magnetic-field-induced spin splitting. When a finite microwave detuning is introduced [$\Delta_d=0.2$, Fig.~\ref{ZsT_all}(c)], the broad high-efficiency regions evolve into more localized structures. The microwave field redistributes the spectral weight among multiple photon-assisted channels, which modifies the resonance conditions and slightly reduces the maximum achievable efficiency.

A more structured behavior appears in the ring--chord geometry [Figs.~\ref{ZsT_all}(b) and (d)] due to the presence of interference-induced transport pathways. In the undriven regime [$\Delta_d=0$, Fig.~\ref{ZsT_all}(b)], the contour map exhibits distinct high-$Z_sT$ regions concentrated around specific combinations of $q$ and $E_z$, indicating that the thermoelectric response becomes highly sensitive to the interference condition. Compared to the pure ring system, the enhanced localization of the high-efficiency regions demonstrates the stronger energy filtering produced by the additional chord channel. Under microwave irradiation [$\Delta_d=0.2$, Fig.~\ref{ZsT_all}(d)], isolated high-$Z_sT$ islands emerge in the large-$q$ and large-$E_z$ regime, where the combined effects of Floquet sidebands and Fano interference significantly improve the spin-selective transport asymmetry. In this regime, the ring--chord geometry attains Max.$,Z_sT \approx 18$, highlighting the cooperative role of interference and photon-assisted processes in optimizing spin caloritronic efficiency.

Overall, the contour analysis demonstrates that the maximum spin thermoelectric performance is governed by a delicate interplay between lead polarization, Zeeman splitting, microwave driving, and quantum interference. While the pure ring geometry provides comparatively broader operating regions, the ring--chord configuration enables sharper and more tunable enhancement of the spin thermoelectric efficiency through interference-assisted energy filtering.

\section{Conclusion}
\label{sec:conclusions}
In conclusion, we have demonstrated that the interplay between microwave-induced Floquet sidebands and geometry-driven Fano interference in a four-quantum-dot ring--chord structure provides an effective mechanism for enhancing both charge and spin thermoelectric performance.
The introduction of an interdot chord gives rise to sharp Fano resonances that strongly suppress thermal transport while preserving pronounced energy asymmetry, resulting in a substantial enhancement of the thermoelectric figure of merit.
In particular, under microwave driving, the charge figure of merit reaches values as high as $ZT \approx 12$ around $\varepsilon_i \simeq 5$, corresponding to thermoelectric efficiencies approaching a significant fraction of the Carnot limit ($\eta/\eta_C \simeq 0.62$) within the linear-response regime. The efficiency--power trade-off analysis demonstrates that quantum interference enhances both the thermoelectric efficiency and output power, with potential experimental realization in quantum-dot and molecular thermoelectric devices through tunable coherent transport pathways.

When ferromagnetic leads and Zeeman splitting are taken into account, the resulting Floquet--Fano interference becomes spin selective, leading to large spin thermopower and a markedly enhanced spin figure of merit.
Quantitatively, compared with the pure ring geometry, the ring--chord structure exhibits stronger interference-assisted enhancement of the spin thermoelectric response, with the maximum value increasing up to Max.$,Z_sT \approx 18$ under microwave irradiation, while the pure ring geometry reaches comparable values only within narrower parameter regimes of lead polarization and Zeeman splitting.
These enhancements are achieved using experimentally accessible control parameters, including moderate magnetic fields, realistic lead polarizations, and microwave amplitudes routinely employed in quantum-dot and mesoscopic transport experiments.
Our findings therefore establish microwave-driven Floquet engineering of Fano interference as a practical and versatile strategy for modulating thermoelectric and spin-caloritronic efficiency in multi--quantum-dot nanostructures, with clear potential for experimental realization.

\bibliography{references.bib}
\bibliographystyle{unsrt}

\end{document}